\documentclass[aps,prb,twocolumn,floatfix,superscriptaddress]{revtex4-2} 
\usepackage[utf8]{inputenc}
\usepackage{graphicx}
\usepackage{bigstrut}
\usepackage{amsmath}
\usepackage{xcolor}
\usepackage{multirow}
\usepackage{hyperref}
\hypersetup{colorlinks=true,allcolors=blue}
\usepackage{comment}
\usepackage{enumitem}
\usepackage{placeins}
\usepackage{mathptmx}
\usepackage{orcidlink}

\definecolor{airforceblue}{rgb}{0.36, 0.54, 0.66}

\begin{document}

\title{Probing out-of-distribution generalization in machine learning for materials}

\author{Kangming Li\orcidlink{0000-0003-4471-8527}}
\email[]{kangming.li@utoronto.ca}
\affiliation{Department of Materials Science and Engineering, University of Toronto, 27 King’s College Cir, Toronto, ON, Canada.}

\author{Andre Niyongabo Rubungo\orcidlink{0000-0003-3608-2039}}
\affiliation{Vertaix, Department of Computer Science, Princeton University, Princeton, NJ, 08544, USA.}

\author{Xiangyun Lei}
\affiliation{Toyota Research Institute, 4440 El Camino Real, Los Altos, California 94022, USA.
}

\author{Daniel Persaud}
\affiliation{Department of Materials Science and Engineering, University of Toronto, 27 King’s College Cir, Toronto, ON, Canada.}

\author{Kamal Choudhary\orcidlink{0000-0001-9737-8074}}
\affiliation{Material Measurement Laboratory, National Institute of Standards and Technology, 100 Bureau Dr, Gaithersburg, MD, USA.}

\author{Brian DeCost\orcidlink{0000-0002-3459-5888}}
\affiliation{Material Measurement Laboratory, National Institute of Standards and Technology, 100 Bureau Dr, Gaithersburg, MD, USA.}

\author{Adji Bousso Dieng\orcidlink{0000-0001-5687-3554}}
\affiliation{Vertaix, Department of Computer Science, Princeton University, Princeton, NJ, 08544, USA.}

\author{Jason Hattrick-Simpers\orcidlink{0000-0003-2937-3188}}
\email[]{jason.hattrick.simpers@utoronto.ca}
\affiliation{Department of Materials Science and Engineering, University of Toronto, 27 King’s College Cir, Toronto, ON, Canada.}
\affiliation{Acceleration Consortium, University of Toronto, 27 King’s College Cir, Toronto, ON, Canada.}
\affiliation{Vector Institute for Artificial Intelligence, 661 University Ave, Toronto, ON, Canada.}
\affiliation{Schwartz Reisman Institute for Technology and Society, 101 College St, Toronto, ON, Canada.}

\begin{abstract}

Scientific machine learning (ML) endeavors to develop generalizable models with broad applicability. However, the assessment of generalizability is often based on heuristics. Here, we demonstrate in the materials science setting that heuristics based evaluations lead to substantially biased conclusions of ML generalizability and benefits of neural scaling. We evaluate generalization performance in over 700 out-of-distribution tasks that features new chemistry or structural symmetry not present in the training data. Surprisingly, good performance is found in most tasks and across various ML models including simple boosted trees. Analysis of the materials representation space reveals that most tasks contain test data that lie in regions well covered by training data, while poorly-performing tasks contain mainly test data outside the training domain. For the latter case, increasing training set size or training time has marginal or even adverse effects on the generalization performance, contrary to what the neural scaling paradigm assumes. Our findings show that most heuristically-defined out-of-distribution tests are not genuinely difficult and evaluate only the ability to interpolate. Evaluating on such tasks rather than the truly challenging ones can lead to an overestimation of generalizability and benefits of scaling.

\end{abstract}

\maketitle

\section{Introduction}

Machine learning (ML) has emerged as an important tool in accelerating scientific discovery~\cite{butler2018machine,carleo2019machine,krenn2022scientific,huang2023central}. This transition to data-driven science is epitomized by the development of scientific ML that seeks to build generalizable models capable of broad applicability~\cite{liu2021towards}. In chemical and materials sciences, recent studies have been focused on developing universal or foundational deep learning models, which are suggested to achieve unprecedented levels of out-of-distribution (OOD) generalizations towards unseen materials that are dissimilar to the training data~\cite{merchant2023scaling,yang2024mattersim,batatia2023foundation,schmidt2022large,deng2023chgnet,chen2022universal}.

However, a critical issue that has been overlooked is the potential biases in selecting OOD tasks to demonstrate generalizability. OOD tasks are often defined based on simple heuristics, which, due to their subjective nature, vary between studies and even lead to contradicting interpretation of generalizability. For instance, the generalization to structures with 5+ elements, despite their omission from training, was used to showcase the emergent capability of deep learning models~\cite{merchant2023scaling}. However, this perspective has been contested with the argument that such generalizations are anticipated from the physical heuristic that interactions in higher-order systems can be inferred from lower-order ones~\cite{li2024efficient,chen2023map,bokas2021unveiling,chen2018database}. This discrepancy underscores a broader lack of agreement and discussion on what constitutes a genuinely challenging OOD task. Indeed, if the OOD test set falls within the training domain, conclusions in the superiority of a state-of-the-art ML architecture and the real benefits of model scaling may pertain only to interpolation capabilities rather than true extrapolation. 

In this work, we conduct a systematic examination of the performance of various ML methods across over 700 OOD tasks within large materials datasets. These tasks are specifically designed to challenge common heuristics based on chemistry or structural symmetry. Our findings indicate that existing models, including those as simple as tree ensembles, exhibit robust generalization across most tasks that feature new chemical or structural groups absent from the training data. Our results highlight that most of these heuristics-based criteria do not constitute truly challenging tasks for ML models. By analyzing the representation spaces of materials, we reveal that test data from well performing tasks largely reside within the training domain, whereas those from poorly performed tasks do not. Furthermore, we find that OOD generalization performance does not necessarily follow traditional scaling laws~\cite{frey2023neural,kaplan2020scaling}. Notably, scaling up training set size or training time leads to marginal improvement or even degradation in the generalization performance for those challenging OOD tasks. These findings indicate that the purported benefits of scaling and emergent generalizability could be considerably overstated, due to domain misidentification driven by human bias that confounds regimes of interpolation and extrapolation.

\section{Results}

\subsection{Evaluation setup}

ML models are usually evaluated for their in-distribution (ID) performance by using random train-test split of the whole dataset. In this work, we assess model performance on OOD tasks, where a specific group of materials serves as the test set while the remainder of the dataset forms the training data. We primarily consider six criteria for defining OOD test data: (1) materials containing element $X$, (2) materials containing any element in the period $X$, (3) materials containing any element in the group $X$, (4) materials of space group $X$, (5) materials of point group $X$, (6) materials of crystal system $X$. We consider these leave-one-$X$-out 
% ($X$ = element, period, group, space group, point  group, crystal system) 
tasks for all possible values of $X$ but exclude those with fewer than 200 test samples. 

Three \textit{ab initio}-derived materials databases have been selected for this evaluation: Joint Automated Repository for Various Integrated Simulations (JARVIS)~\cite{choudhary2020joint,Wines2023}, Materials Project (MP)~\cite{jain2013commentary}, and the Open Quantum Materials Database (OQMD)~\cite{Saal2013}. These databases have different data distributions, hence ensuring a robust and generalized conclusion of OOD performance. The combination of datasets and grouping criteria leads to over 700 OOD tasks. OOD performance is evaluated by training a representative set of ML models, including (1) random forest (RF)~\cite{breiman2001random} and XGBoost (XGB)~\cite{xgboost} models with Matminer descriptors~\cite{ward2018matminer}. (2) single neural network with Gaussian multipole (GMP) expansion on electron density~\cite{lei2022universal} (3) atomistic line graph neural network (ALIGNN)~\cite{choudhary2021atomistic}, and (4) the large language model (LLM) based LLM-Prop with crystal text descriptions~\cite{rubungo2023llm}. It therefore covers not only common ML architectures from lightweight tree ensembles and neural networks to graph neural networks and transformer-based large language models, but also distinct input representations from human-devised descriptors and force-field-like features to crystal graphs and text.

\subsection{Chemistry-based OOD generalization}

In the following, our analysis primarily focuses on formation energy data, given its fundamental importance in materials science. Table~\ref{tab:ID-performance} shows the in distribution (ID) performance obtained from a random 8:2 train-test split of the whole dataset, providing a baseline for comparison with OOD performance. We utilize two complementary performance metrics: mean absolute error (MAE) and coefficient of determination ($R^2$). While MAE measures the expected error on the original physical scale, its interpretation can be scale-dependent and less intuitive for assessing the goodness of predictions. In contrast, $R^2$ is a dimensionless accuracy measure with a value range between 1 (for a perfect model) and infinitely negative (for arbitrarily bad models) and can be conveniently compared across different OOD test sets.

\begin{table}
\caption{In-distribution performance for the formation energy prediction. Models are arranged in the ascending order of MAEs of the MP dataset from left to right. Best performance is highlighted in bold. The MP, JARVIS, and OQMD datasets contain 146k, 76k, and 1M entries, respectively.
}
\label{tab:ID-performance}
\begin{ruledtabular}
\begin{tabular}{ccccccc}
Metric               & Dataset & ALIGNN & GMP   & LLM-Prop & XGB   & RF    \\
\hline
\multirow{3}{*}{MAE (eV/at)} & MP      & \textbf{0.033}  & 0.052 & 0.063    & 0.078 & 0.090 \\
                     & JARVIS  & \textbf{0.036}  & 0.081 & 0.068    & 0.074 & 0.099 \\
                     & OQMD    & \textbf{0.020}  & 0.038 & 0.045    & 0.070 & 0.065 \\
\hline
\multirow{3}{*}{$R^2$}  & MP      & \textbf{0.996}  & 0.992 & 0.981    & 0.979 & 0.970 \\
                     & JARVIS  & \textbf{0.995}  & 0.985 & 0.982    & 0.981 & 0.968 \\
                     & OQMD    & \textbf{0.998}  & 0.995 & 0.995    & 0.987 & 0.985
\end{tabular}
\end{ruledtabular}
\end{table}

\begin{figure*}
    \centering
    \includegraphics[width=\linewidth]{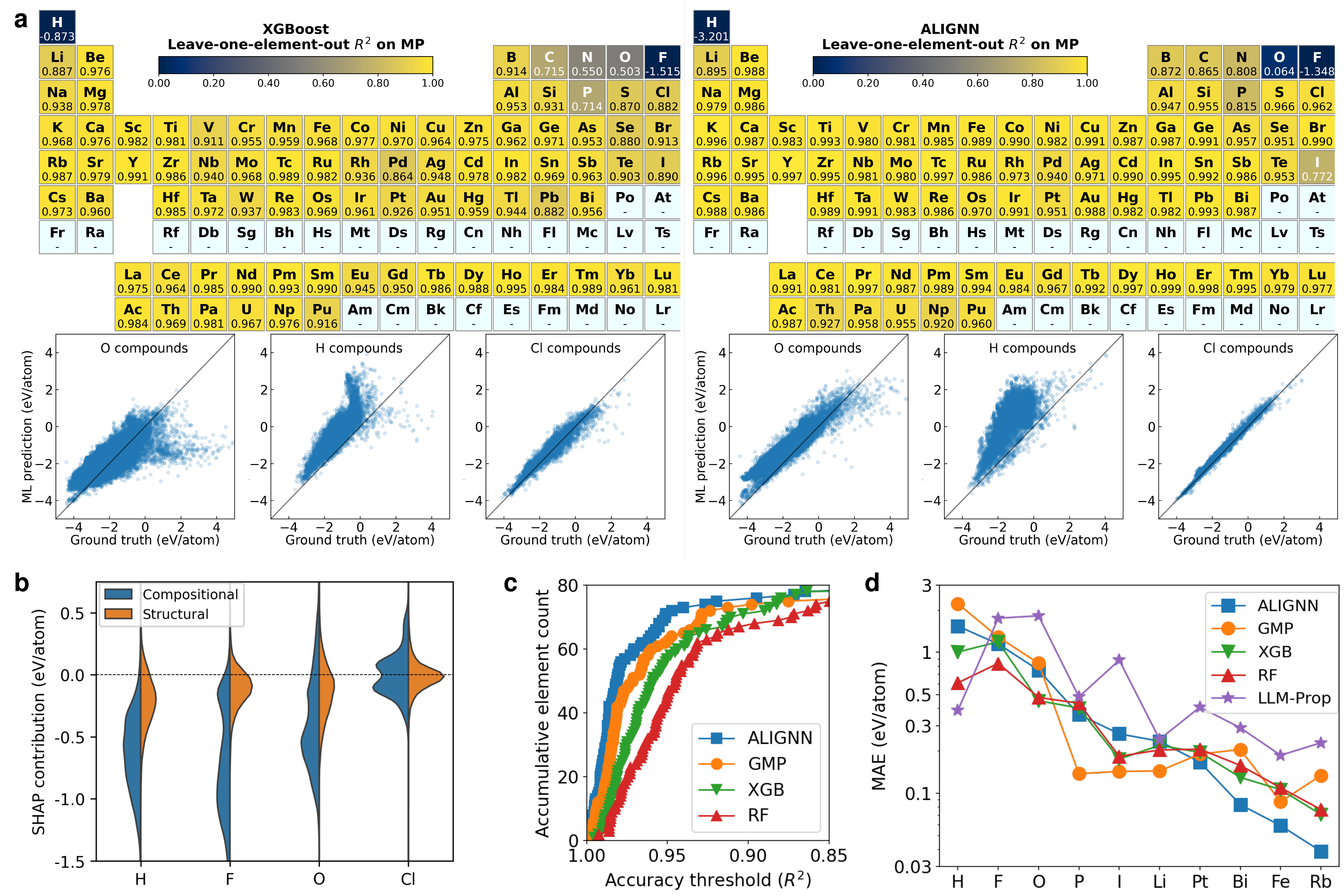}
    \caption{Leave-one-element-out performance on Materials Project. (a) Leave-one-element-out performance across the periodic table for the XGB and ALIGNN models. Despite the presence of negative $R^2$ for H and F, the colorbar range is bounded between 0 and 1 for better visibility. The OOD tests are performed only for elements with more than 200 data. Results for other models and datasets are provided in Supplemental Material. The periodic tables are plotted using pymatviz~\cite{riebesell_pymatviz_2022}. (b) Violin plots of SHAP contributions from compositional and structural features for selected left-out elements. (c) Number of leave-one-element-out tasks with $R^2$ higher than a given threshold as a function of threshold.
    % The horizontal dashed line indicates the total number of tasks. 
    LLM-Prop is not included due to the large number of tasks and high training cost. (d) Comparison of performance between ML models for selected leave-one-element-out tasks. 
    }
    \label{fig1a}
\end{figure*}

\begin{comment}
    There are two problems if using the MAE_OOD to 
    MAE_ID: (1) the training set size for MAE_ID: how to (2) We still want to see the absolute performance. A poor model won't have a strong performance degradation. We can provide this result in the supplemental material though.
\end{comment}

Figure~\ref{fig1a}(a) illustrates the leave-one-element-out generalization performance of the ALIGNN and XGB models on the MP dataset, demonstrating robust OOD performance across much of the periodic table. Notably, 85~\% of the tasks for the ALIGNN model and 68~\% for the simpler XGB model achieved $R^2$ scores above 0.95. This broad generalization capability is surprising, given that the training set does not contain any information regarding the bonding between the left-out element and other atoms. Our results suggest that effective OOD generalization across chemistry may be more achievable than previously assumed, for both low- and high-capacity models.

Tasks with low $R^2$ scores are mainly associated with nonmetals such as H, F, O. Nonetheless, ML models may still prove useful in ranking materials even in the worst performing tasks. As shown in the parity plots, the ALIGNN model systematically overestimates the formation energies of H compounds (or O compounds), with a Pearson correlation coefficient of 0.7 (or 0.9). The poor $R^2$ is therefore a result of systematic biases in the OOD predictions. Such bias is also shared among other ML models: all ML models considered here overestimate formation energies of H and F compounds; the three deep learning models tend to overestimate formation energies of O compounds, while tree models tend to overestimate low-energy and underestimate high-energy O compounds. Similar systematic biases, which can be addressed by simple linear corrections, have also been reported elsewhere for OOD property predictions~\cite{deng2024overcoming,choudhary2023can}. Future investigation to understand and mitigate these biases can be useful for developing chemically transferable models.

The causes behind the poor OOD performance for nonmetals, particularly H, F, and O, have been examined. Initially, we considered training set size as a potential factor, but this was ruled out based on the weak correlation between training set size and OOD performance, evidenced by a Spearman's rank correlation coefficient of -0.2. Despite investigating various element attributes, none could fully account for the $R^2$ trends. For example, while elements with low $R^2$ scores tend to be more electronegative or positioned at the boundary of the periodic table, exceptions such as S, Cl, Br (which are electronegative) and Cs, Bi (positioned near the corner of the periodic table) exhibit high ALIGNN $R^2$ scores above 0.96, suggesting other influencing factors.

The biases leading to this poor OOD performance could be of either compositional or structural origin. The former is related to chemical dissimilarity between elements, whereas the latter refers to the failure mode where the test set associated with a given element contains materials that are structurally distinct from the training set. We propose a SHAP-based~\cite{lundberg2020local2global} (SHapley Additive exPlanations) method to identify the sources of biases. This involves training a second model to correct the model for each leave-one-element-out task and evaluating the contributions from compositional and structural features to the corrections (see Method). Fig.~\ref{fig1a}(b) shows the violin plots of the feature contributions from the XGB model for the test data in selected leave-one-element-out tasks. Compared to the well-performing case of Cl, the compositional contributions in the cases of H, F, and O are much stronger in magnitude than structural contributions. The predominately negative compositional contributions are consistent with the corrections needed for the overestimation shown in the parity plots. This analysis therefore reveals that the poor OOD performance is mainly of chemical origin, namely the chemistry-property relationship is drastically different and cannot be readily transferred from other chemical systems.

\begin{figure*}
    \centering
    \includegraphics[width=\linewidth]{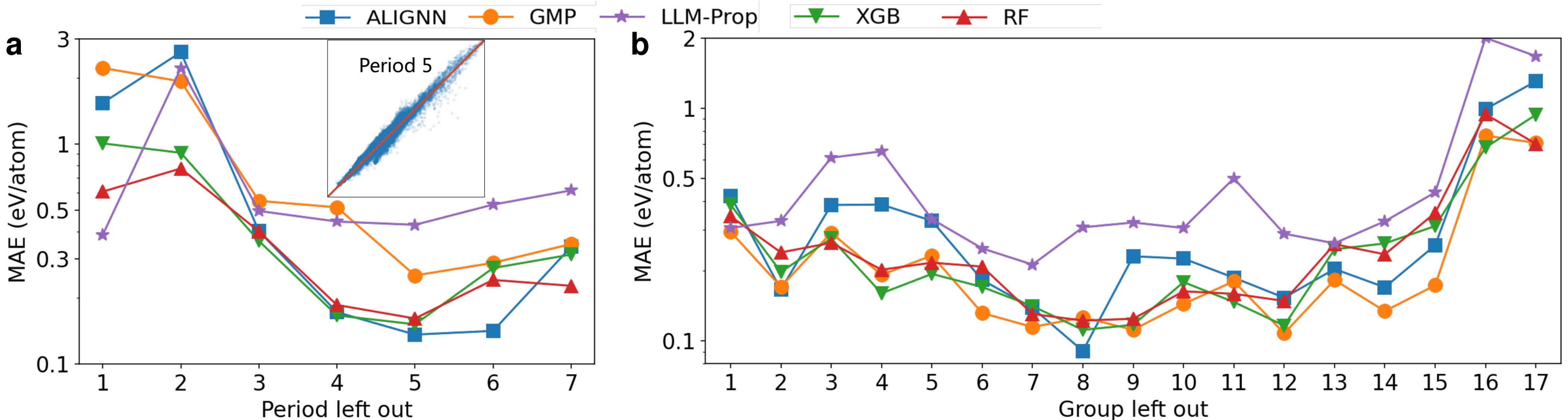}
    \caption{Leave-one-period-out and leave-one-group-out performance on Materials Project. The inset in (a) shows the parity plot for the ALIGNN predictions for the period 5 (axis range: -5 eV/atom to 5 eV/atom).
    }
    \label{fig1b}
\end{figure*}

To compare the overall performance between different models, we count for each model the number of leave-one-element-out tasks with an accuracy better than a given threshold. As shown in Fig.~\ref{fig1a}(c), models with better ID performance have a better overall performance in the leave-one-element-out tasks. However, this correlation does not imply uniform superiority of any model across all tasks. Figure~\ref{fig1a}(d) details performance comparisons in 10 representative tasks, highlighting nuanced outcomes. In the most challenging tasks involving H, F, and O, the RF model, despite its lower ID performance, ranks as the best or second best. Conversely, for moderately challenging tasks (P, I, and Li), the GMP model outperforms others. Meanwhile, in less demanding tasks (Bi, Fe, and Rb) where MAEs are below 0.1 eV/atom, the ALIGNN model achieves a much lower MAE than the other models. 

Interestingly, the LLM-Prop model, despite its good ID performance, is the worst performing model in most of the leave-one-element-out tasks. A possible cause might be the lack of inductive bias related to the element similarity, which is encoded in the representation schemes of other models. For instance, ALIGNN and tree ensembles represent atoms by basic attributes such as electronegativity, period, and group. In this way, the models can properly recognize the correlation between the left-out element and the other elements in the training set. By contrast, language models treat each element as a text token in a manner equivalent to one-hot encoding. Element similarity may not be encoded in the initial text embedding and is not guaranteed to be learned for the left-out element during the training. A potential direction to improve chemical transferability of LLMs is therefore to include additional chemical information in text descriptions during pretraining or fine-tuning. Alternatively, language models specialized for materials and chemistry applications might explore custom tokenization schemes for elemental symbols and chemical formulae that directly incorporate aspects of chemical similarity.

It is worth noting that many representation schemes that use one-hot encoding of elements or element-specific functions~\cite{behler2007generalized} are in principle incapable of any generalization to new chemistry. By contrast, the good generalization across chemistry shown in Fig.~\ref{fig1a} highlights the importance of describing elements with a transferable representation. In this way, an element and its compounds can be seen as lying within the interpolation region in the chemical space of the training data in the leave-one-element-out tasks. 

Appropriate representations can achieve chemical transferability even for small training sets with limited chemical diversity. To prove this point, we create a dataset from OQMD by restricting the chemistry to six elements from Cr to Cu. The leave-one-element-out $R^2$ scores of the XGB model within this dataset (about 1200 structures) range between 0.87 (for Cu) to 0.93 (for Fe), comparable to the in-distribution $R^2$ score of 0.93, demonstrating that generalization across chemistry does not necessarily require a big model or large dataset.

% Indeed, as there is no significant OOD performance degradation for a large portion of the periodic table, the overall OOD performance ranking of ML models
% % , evaluated by the number of OOD tasks with accuracy above a given threshold, 
% follows the same order as the ID performance ranking shown in Table~\ref{tab:ID-performance}. {\bf The only exception here is the LLM-Prop model. Token limits, instability of the convergence.}

Fig.~\ref{fig1b}(a) shows the more challenging OOD tasks where an entire period of elements is excluded from the training set. The worst OOD performance is found for the first period, which contains H, and the second period, which contains many nonmetals with poor leave-one-element-out performance. Overall, the OOD performance tends to be better for the periods in the middle of the periodic table, where ML models can learn from the adjacent periods. For the OOD tasks of the fourth to sixth periods, the $R^2$ of the ALIGNN model can achieve a $R^2$ score of 0.95 to 0.97, which is surprisingly good given the number of elements excluded from the training set. Interestingly, the XGB and RF models perform similarly or even better than deep learning models in most of the tasks, highlighting the robustness of tree ensembles in challenging generalization tasks.

Similar conclusions are also found in the leave-one-group-out performance shown in Fig.~\ref{fig1b}(b). First, the OOD performance also tends to be better for the groups in the middle of the periodic table. Despite its superior leave-one-element-out performance for elements in groups 3 to 5, the ALIGNN model is outperformed by the tree ensembles in these leave-one-group-out tasks. By contrast, the GMP model demonstrates the best or close-to-best performance in these tasks. The difference in the model performance ranking in the leave-one-element/period/group-out tasks indicates that these ML models have distinct advantages in different types of OOD tasks. Future investigation in understanding their respective advantages can be beneficial for designing models with better generalizability. 

In the JARVIS and OQMD datasets, we also find good generalization performance across the majority of the periodic table and poor performance for a few cases of nonmetal elements. Therefore, the trend shown in Fig.~\ref{fig1a} and~\ref{fig1b} is not a unique artifact of the specific biases in data distribution or computational settings of the MP dataset, nor of the training set size which will be discussed later. Instead, it reveals that the existing featurization schemes can enable transferability between most of the chemical systems, though they may still fail to capture the essential characteristics for certain nonmetals, which leaves rooms for future improvement in materials featurization.

\subsection{Structure-based OOD generalization}

\begin{figure*}
    \centering
    \includegraphics[width=\linewidth]{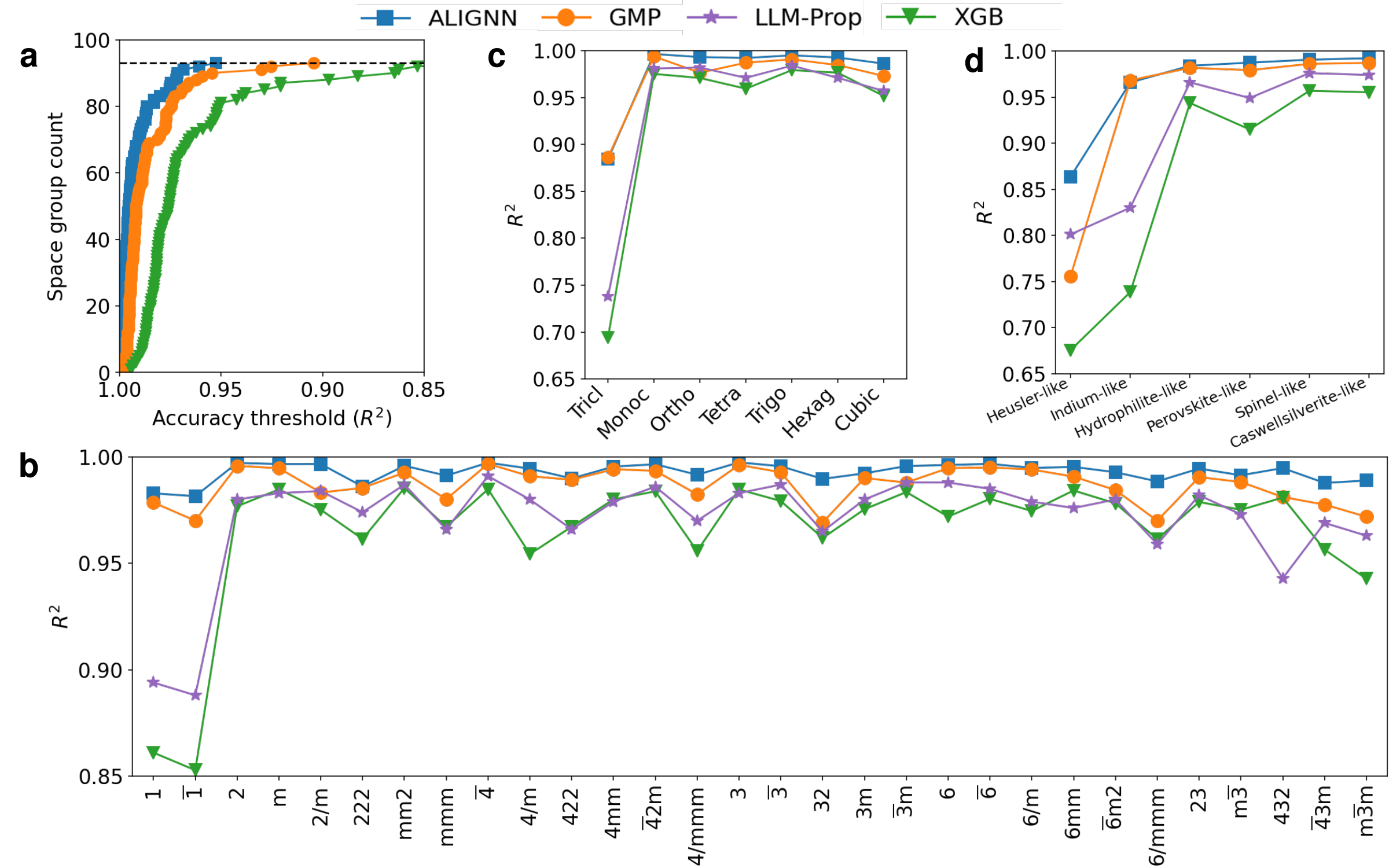}
    \caption{
    Structure-based OOD performance on Materials Project. (a) 
    Number of leave one space group out tasks with $R^2$ higher than a given threshold as a function of threshold. The horizontal dashed line indicates the total number of tasks. LLM-Prop is not examined due to the large number of tasks and high training cost. (b) Leave one crystal system out performance in the ascending order of symmetry from left to right. (c) Leave one prototype-like out performance. Structures with fingerprints similar to a prototype are grouped together. (d) Leave one point group out performance in the ascending order of symmetry from left to right. 
    }
    \label{fig2}
\end{figure*}

Structures are usually grouped by their crystallographic symmetries. Here we consider three basic symmetry attributes for structural grouping: space groups, point groups, and crystal systems, with each being a subcategory of the next. The structure-based OOD performance on the MP formation energy dataset are shown in Fig.~\ref{fig2}.

The OOD performance of the structure-based tasks is much better than that of the chemistry-based OOD tasks. Fig.~\ref{fig2}(a) shows the overall OOD performance on all of the leave one space group out tasks. As the best performing model, ALIGNN achieves an $R^2$ of above 0.95 in all of the tasks with 88~\% of them having an $R^2$ score above 0.98. The GMP model is the second best performing one followed by the XGB model. The LLM-Prop model is not examined here due to the high training cost required for a large number of tasks, but it is included for the leave one point group out and leave one crystal system out tasks in Fig.~\ref{fig2}(b) and (c). All models demonstrate a good prediction accuracy with an $R^2$ of above 0.95 (or a MAE of below 0.15 eV/atom) in most of these tasks. The only cases with relatively poor performance are associated with the triclinic crystal system (and its subcategory point groups 1 and $\overline{1}$), which has minimal symmetry constraint and is structurally more diverse, posing greater challenge when generalized from structures with higher symmetry.

Our results suggest that symmetry attributes, which are often used to explain generalization performance or devise active learning algorithms~\cite{li2022critical,zhang2023entropy,schrier2023pursuit,schmidt2022large,goodall2022rapid}, may not be effective in finding data truly dissimilar from the training set. Indeed, two symmetrically different crystals can be structurally very similar. For example, a slight distortion of a cubic lattice can transform the structure into another lattice, as in the famous cubic-to-tetragonal Martensitic transformation in steels. Therefore, the training set still contains similar structures even if all cubic structures are removed. To mitigate this information leakage and make the generalization task more challenging, we create new test sets by using the fuzzy prototype-matching algorithm implemented in the robocrystallographer package~\cite{ganose2019robocrystallographer} to group prototype-like structures. Fig.~\ref{fig2}(d) shows the OOD performance on a few leave one prototype-like out tasks. Interestingly, the OOD performance in these tasks is still good, especially for deep learning models. The ALIGNN and GMP models achieve $R^2$ scores above 0.95 in all tasks except for Heusler-like structures, for which the ALIGNN model can still achieve an $R^2$ of 0.86 and an MAE of 0.08 eV/atom. The XGB model still perform reasonably well in these tasks, though its MAEs are 100~\% to 200~\% higher than those of the ALIGNN model. 

Overall, the ranking of model performance in structure-based OOD tasks closely follows the ID performance ranking, in contrast to the situation in chemistry-based OOD tasks. In particular, the ALIGNN model systematically outperforms other models in all the structure-based tasks, highlighting the effectiveness of graph neural networks in capturing complex structural patterns. In addition, the LLM-Prop model also exhibits much better generalization across structural groups than across chemical groups.

\subsection{Inspecting materials representations}

\begin{figure*}
    \centering
    \includegraphics[width=\linewidth]{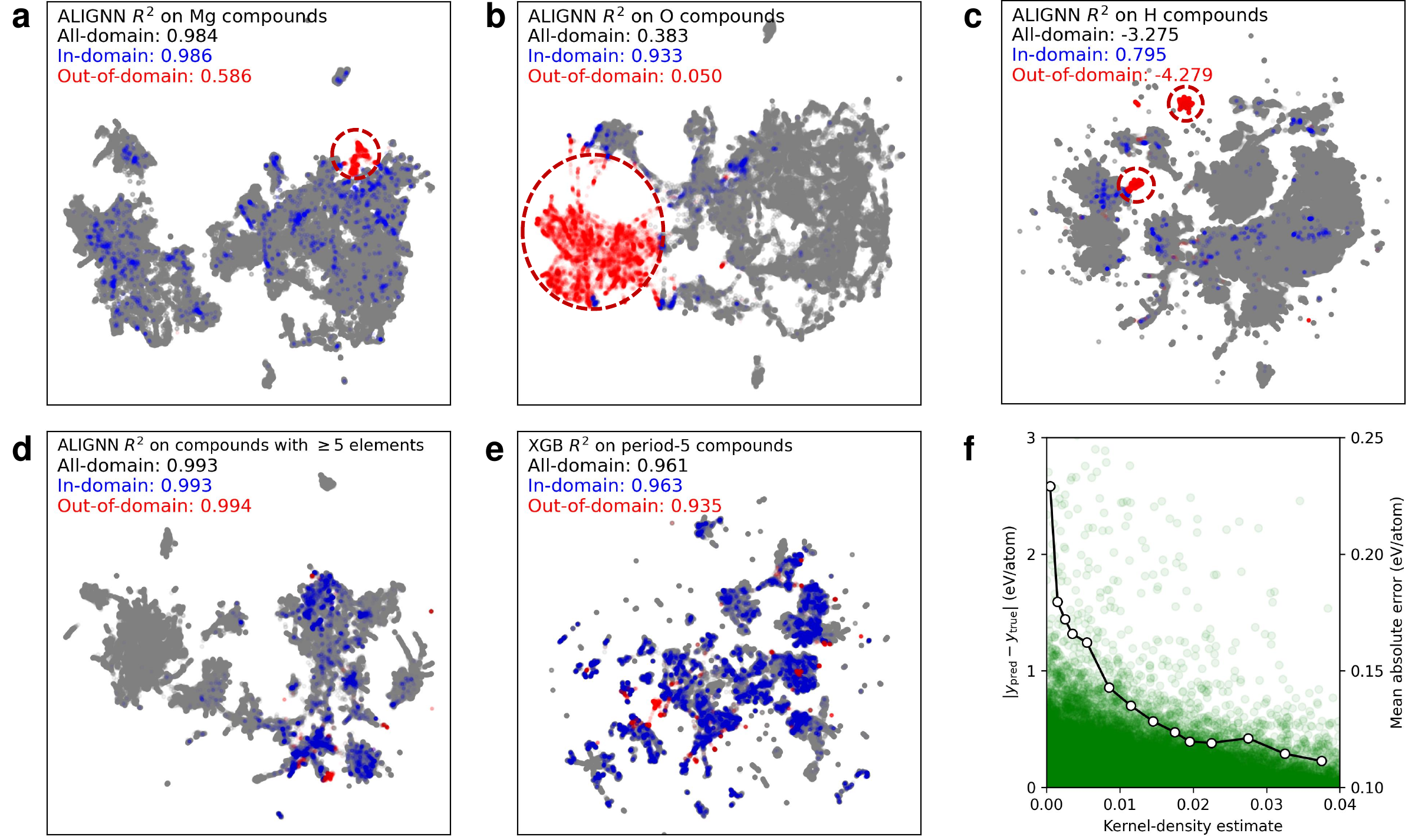}
    \caption{UMAP two-dimensional projection of materials representations for domain identification. (a-b) UMAP plots of ALIGNN embeddings learned from the leave-Mg-out and leave-O-out tasks in JARVIS. (c) UMAP plot of ALIGNN embeddings learned from the leave-H-out task in OQMD. (d) UMAP plot of ALIGNN embeddings learned by leaving out structures with 5 or more elements in MP. (e) UMAP plot of the XGB descriptors for the leave-period-5-out task in MP. (f) Absolute errors (left Y axis) of test data as functions of kernel-density estimates of training data for the UMAP plot of (e); the solid line denotes the MAEs (right Y axis) for different density intervals. In all UMAP plots, the training data, in-domain test data, and out-of-domain test data are marked in grey, blue, and red, respectively; clusters of out-of-domain test data are circled out in (a-c); the $R^2$ scores are indicated for the in-domain, out-of-domain, and all-domain test data.
    }    
    \label{fig4}
\end{figure*}

The systematic investigation of various OOD tasks reveals that in most of the cases, generalization beyond the distribution of the training data is not challenging. It therefore seems to suggest that ML models are good at extrapolating across chemical and structural groups, in contrast to the recent findings that ML models generalize poorly even between different versions of the same databases~\cite{li2022critical,li2023exploiting}. To reconcile this apparent contradiction, we propose to inspect materials not from the perspective of a single human-defined attribute (e.g., element identity), but from the high-dimensional representation space of materials. 
% Furthermore, we propose to distinguish between the concepts of distribution and domain. A test set is considered out-of-distribution if the statistical distribution of one or more attributes is different from the training set, whereas a test set is considered out-of-domain if the test data are outside the domain of the training set in the high-dimensional representation space. 
Furthermore, we propose to define and distinguish two types of OOD. A test set is considered statistically OOD if the statistical distribution of one or more attributes differs from the training set, whereas a test set is considered representationally OOD if the test data are outside the region of the training set when viewed from the high-dimensional representation space. With this definition, all the OOD tasks discussed previously are statistically OOD (which will be simply referred to as OOD) but not necessarily representationally OOD.

Uniform Manifold Approximation and Projection (UMAP)~\cite{umap} is used to project the high-dimensional materials representations to a two-dimensional plane. For deep learning models, materials representations (embeddings) are automatically learned during the training; embeddings of new materials can then be derived by using the trained models. For each OOD task, we apply a Gaussian kernel to the training data and evaluate the kernel density estimate of the training data for every test data point in the UMAP embedding space. Test data points with high/low density estimates are considered as representationally ID/OOD data.

\begin{figure*}
    \centering
    \includegraphics[width=\linewidth]{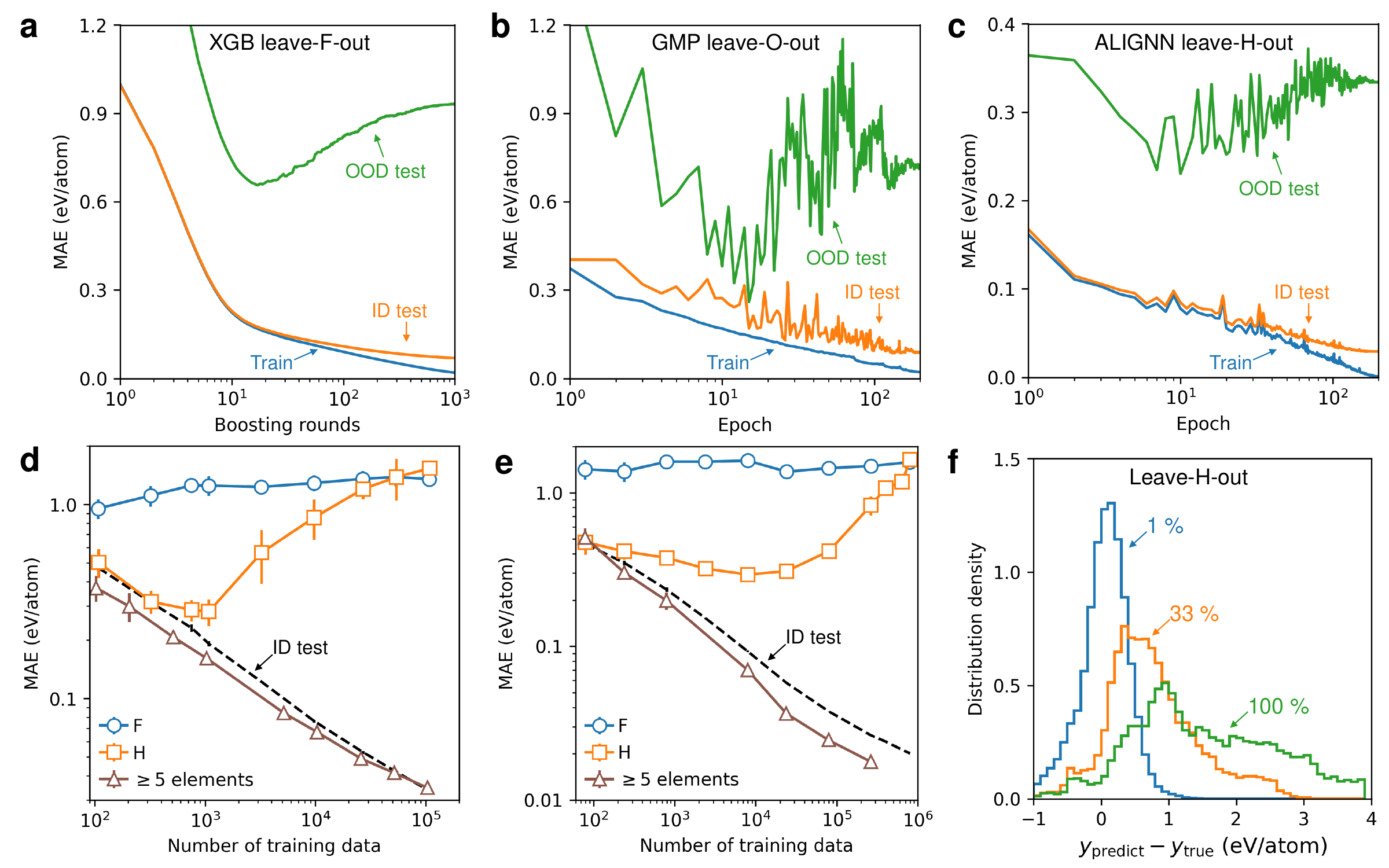}
    \caption{Effects of training time and training set size on performance. (a-c) Training loss, ID test loss, and OOD test loss as functions of training round/epoch on JARVIS. Models and left-out elements are indicated in the top of the plots. 
    (d-e) ALIGNN performance on Materials Project and OQMD, respectively. Solid lines with markers denotes the MAEs as functions of training set size for 3 representative OOD tasks: leave F out, leave H out, and leave out structures with $\geq$ 5 elements. Dash line marks the ID MAE as the baseline for comparison. 
    (f) Distribution of ALIGNN prediction errors on H-containing structures in the leave-H-out task in OQMD. Results for the models trained on 1~\%, 33~\%, and all of the structures without H are shown. Similar analysis for other tasks and ML models is provided in Supplemental Material.
    }    
    \label{fig3}
\end{figure*}

Fig.~\ref{fig4}(a-b) show the UMAP plots of ALIGNN embeddings learned from the leave-Mg-out and leave-O-out tasks in the JARVIS dataset. In the former case, 76~\% of Mg-containing structures are well covered by the training data (i.e., with an embedding space probability density above the threshold), whereas 24~\% of Mg-containing structures are representationally OOD with a much lower $R^2$ score. In the latter task, 87~\% of O-containing structures are representationally OOD with an $R^2$ close to 0, whereas there are still some in-domain O-containing structures with an $R^2$ above 0.9. Fig.~\ref{fig4}(c) shows another example from the leave-H-out task in the OQMD dataset, where kernel-density estimates provide again good resolution in separating representationally ID and OOD test data: 13~\% of H-containing structures are identified as within the training domain with an $R^2$ score of 0.8. The identification of poorly predicted test data in well-performing OOD tasks and well-predicted test data in poorly-performing OOD tasks demonstrates the effectiveness of the proposed method for classifying representational domains of OOD samples. 

Representational domain identification is important for interpreting generalization capabilities. Fig.~\ref{fig4}(d) shows the case where the test set contains structures with 5 or more elements. It has been used as an example to demonstrate the emergent ability of deep learning models~\cite{merchant2023scaling}. This perspective was contested as the same capability was also found in simple boosted trees~\cite{li2024efficient}. In fact, our UMAP analysis shows that 92~\% the test data are identified as lying within the representational domain of the training data, therefore this test only reflects the interpolation capacity of ML models. This example highlights not only the risk of misinterpreting generalization capabilities, but also the need for designing more challenging tests that can truly demonstrate state-of-the-art.

Domain identification can be achieved not only with the learned embeddings but also with the descriptors used in tree ensembles. An example is shown in Fig.~\ref{fig4}(e), where the test data are structures that contain any element in the fifth period. Based on kernel-density estimates, 93~\% of the test data are identified as representationally ID data, thereby explaining the good generalization performance for this seemingly difficult task. 

It is worth noting that density estimates do not perfectly correlate with prediction errors, as shown in Fig.~\ref{fig4}(f). However, from a statistical perspective there is a higher proportion of test data with high prediction errors in the low-density region than in the high-density region. Indeed, when calculating the MAE for test data in different density intervals, we find that the MAE tends to decrease with increasing density estimate (dotted line). The density estimate can still enable us to statistically distinguish between the well-predicted in-domain data and poorly-predicted out-of-domain data from an out-of-distribution test set. Future work will be needed to improve domain identification and uncertainty measures with better correlation with OOD prediction errors.

\subsection{Learning curves for out-of-distribution generalization}

Domain identification in the materials representation space is important in correctly interpreting the benefits of the neural scaling strategy for materials discovery applications. This strategy is based on the so-called neural scaling laws, where test errors can be consistently reduced by increasing model size, dataset size, and the amount of compute used for training deep learning models. Recent work on deep learning models for materials follows this rationale to train foundational or universal neural networks
~\cite{merchant2023scaling,deng2023chgnet,riebesell2023matbench,batatia2023foundation,schmidt2022large}. However, neural scaling laws have been empirically tested only for the ID generalization~\cite{frey2023neural,faber2017prediction}. Here we perform an extensive examination on various OOD tasks and, intriguingly, find distinct scaling effects among the OOD tasks. Specifically, neural scaling laws are found to be valid only for representationally ID generalization, while the scaling effects are marginally beneficial or even adverse for representationally OOD generalization.

Fig.~\ref{fig3}(a-c) shows the learning curves of MAE versus training time in the leave-one-element-out tasks for F, O, and H. The training and the ID set contain 80~\% and 20~\% of the structures without the left-out element, whereas the OOD test set contains the structures with the left-out element. For the leave-F-out task, the training and ID test MAEs decrease with increasing training time of the XGB model. However, the OOD MAE decreases only during the initial 20 training rounds, then continuously increases with more training.

A similar pattern is observed in the leave-O-out and leave-H-out tasks. In fact, for representationally OOD tasks, the test MAE tend to increase or remain constant with increasing training time beyond around 20 epochs. However, for OOD tasks that are representationally ID, the OOD MAE has a tendency similar to the ID MAE, namely the error decreases with increasing training time. As discussed in Fig.~\ref{fig4}, a key distinction between the easy and difficult OOD tasks is whether the test data lies outside the training domain in the representation space. Thus, our findings indicate that the response to scaling up training efforts may vary significantly based on whether the test data falls inside or outside the training domain.

% To mitigate the performance degradation, it is advantageous to implement early-stopping schemes by tuning on ID data. As illustrated in Figure~\ref{fig3}(a), the learning curves for the XGB model’s training and ID test MAEs exhibit a change in slope between 10 and 20 training rounds. Beyond this point, there is weak reduction in ID test MAE but a notable deterioration in OOD performance. This pattern, observed across various tasks, suggests that the point of slope change in an ID validation set can serve as an effective early-stopping criterion to enhance the generalizability and prevent overfitting to the training domain. However, this does not apply to deep learning models, for which identifying similar characteristics are worth further investigation. 

Effects of training set size are also investigated. Take the leave-F-out task for example, models are trained with different sizes of the training sets randomly sampled from 80~\% of the structures without F, and evaluate performance on the held-out ID test set (20~\% of the structures without F) and the OOD test set (structures with F). 

Fig.~\ref{fig3}(d) shows the ALIGNN learning curves for the ID and OOD performance for 3 representative tasks in Materials Project. As the ID learning curves for different OOD tasks are found to be almost the same, we show only one learning curve for the ID performance (dashed line) for simplicity. The ID performance consistently improves with increasing training set size, which is however not always the case for OOD performance. For the leave-F-out task, increasing the training set size from $10^2$ to $10^5$ only further degrades the already bad OOD performance. For the leave-H-out task, we find a ``V" shaped learning curve for the performance on H-containing structures, where the MAE first decreases and then increases with increasing training set size. This violation of neural scaling laws is also found in other representationally OOD tasks (Supplemental Material). By contrast, we find a consistent decrease in MAE with increasing training set size, for the case where the test data are structures with 5 or more elements. This accordance with neural scaling laws is also found in other easy OOD test sets that contain mainly representationally ID data (Supplemental Material). 

To avoid establishing our conclusion based on a single dataset, we also perform the same experiments on the OQMD dataset, which is about an order of magnitude larger than the Materials Project dataset. As shown in Fig.~\ref{fig3}(e), we find the same trend where increasing training set size only leads to marginal improvement or even degradation for test data that are outside the training domain. It should be noted that the marginal improvement is not an artifact of the limited model capacity, considering the continuous decrease in the ID test MAE with increasing training set size. In addition, similar learning curves are also found for other ML models and the JARVIS dataset (Supplemental Material).

It is particularly striking to observe the 5-fold increase in the OOD MAE of the leave-H-out task, when the training set size is increased from $10^4$ to $10^6$. Fig.~\ref{fig3}(f) shows the distributions of the corresponding prediction errors on H-containing structures. When trained on only 1~\% of structures without H, the error distribution is centered around zero with a relatively narrow spread. With increasing training set size, however, the errors shift towards positive values and exhibit a larger spread. This systematic overestimation is a reflection of biases in models that are enhanced by training on more data from a given domain. The leave-H-out task can therefore serve as an example of losing generalizability due to an overfitting to the training domain, for which future work is needed to identify its root cause and mitigation solutions. 

Our results challenge previous interpretations of generalizability. Figure~\ref{fig3}(d-e) include the learning curves for the scenario where the test set comprises structures with five or more elements, while the training set includes those with fewer than five. Contrary to prior claims suggesting such generalization is an emergent capability enabled by neural scaling~\cite{merchant2023scaling}, our analysis indicates that this may merely reflect in-domain generalization. According to the literature, a sudden improvement in performance above a certain threshold is required to classify an ability as emergent~\cite{wei2022emergent}; however, this criterion is not met in the learning curve. Additionally, the OOD MAE being lower than the ID test MAE further supports our findings from Figure~\ref{fig4}(e) that the test data lie well within the representational training domain. Thus, the observed effective generalization in these cases does not demonstrate the models’ true generalizability but rather the interpolation within the representational training domain, where errors decrease as the test data become increasingly well-covered by the training data.

\begin{comment}

Highlight for out-of-domain samples. 

The motivation here is to show that the characteristics is very different for out-of-domain (OODo)

comment on the the training set size here. No strong correlation. Then say increasing the training set size does not solve the issue.

Explain why: focus on the hard and the easy one

Should we consider them as in-distribution or out-of-distribution? We have seen that some are quite different from the other OOD tasks. 

% We should apply multiple models (mixture of experts, depending on the domain). 
\end{comment}

\section{Discussion}
% We have demonstrated through extensive out-of-distribution examinations that existing models exhibit remarkable generalization capabilities to new chemical or structural groups outside the training set. We explain these observations by inspecting the less intuitive representation space of materials. We show that many out-of-distribution test sets are actually within the domain of training data, whereas only a handful of challenging out-of-distribution tasks contain lots of test data outside the training domain. We find that the performance on out-of-domain tasks does not follow the conventional neural scaling laws. We suggest that the benefits of scaling for out-of-domain generalization may be overstated or misinterpreted.

% Our findings of good generalization certainly demonstrate the predictive power of ML methods, but it also reveals that most tasks commonly considered difficult based on human intuitions are not truly challenging. Choosing such tasks to demonstrate the state-of-the-art performance of ML models therefore reflects mainly interpolation performance but does not truly represent the out-of-domain generalizability. We believe future discussion in what constitutes a genuinely demanding out-of-domain task or challenging splits would create more convincing test cases, with which we can identify, understand, and bridge the gaps towards truly generalizable ML models. 

% {\bf I think here heuristically defined tasks are similar to the initial version of the Turing test. Existing ML models largely pass that test and more challenging tests are needed.}

Our extensive OOD examinations show that existing models exhibit remarkable generalization capabilities to new chemical or structural groups outside the training set. We explain these observations by inspecting materials representation, revealing that many OOD test sets are actually within the representational domain of training data, whereas only a handful of challenging OOD tasks contain significant amounts of test data outside the representational training domain. Intriguingly, the performance on these representationally OOD tasks does not adhere to conventional neural scaling laws, suggesting that the benefits of scaling for OOD generalization may be overstated or misinterpreted.

Our findings underscore the predictive power of ML but also reveal that many OOD tasks commonly considered difficult based on intuitions are not truly challenging for state-of-the-art models. This situation is reminiscent of the Turing test, which was believed to be a golden standard for proving machine intelligence but is now considered insufficient for benchmarking the latest language models~\cite{biever2023chatgpt}. The heuristically defined OOD tasks are akin to this early benchmark, where existing models largely pass yet more challenging tests are necessary to push the boundaries of what is considered state-of-the-art. Choosing such tasks reflects mainly interpolation capabilities of ML models but does not truly represent out-of-distribution generalizability. We advocate for more attention and discussion on what constitutes a genuinely demanding OOD task. This will pave the way for creating more convincing test scenarios, with which we can identify, understand, and bridge the gaps towards truly generalizable ML models.

\begin{comment}

While we adopt this definition of OOD, we consider it is not a good measure, because it is based on human intuition, rather than more objective metrics.

Our systematic evaluation is useful to down-select a much smaller but representative set of OOD tasks for benchmarking. 
Future work can focus on some representative cases.

Generalize better than we think. So we should reexamine the application domain which would be much larger than our heuristic expectation (as evaluated in this work~\cite{batatia2023foundation})

However, the better-than-expected level of generalizability is after all not that surprising. This is just beyond our original bias in domain identification. 
% This reveals a type of anthropogenic biases 
As we have domonstrated here, they do not really do OOD. OOD detection is another issue.
Moving forwards, the community should use stricter criterion, and pay attention to attribute the superior performance to e.g. model architecture or the dataset size.

In practice, leave-one-element-out is rare.
% how to identify the hard-to-generalize grouping. This kind of adversial attach would help understand.

\end{comment}

\section{Methods}

% Tree methods such as RF and XGB are more interpretable and orders of magnitude faster than deep learning models and often used as baselines. GMP models 
% representative methods that use distinct materials representation and ML architectures

% \begin{table*}[]
% \caption{}
% \label{tab:my-table}
% \begin{ruledtabular}
% \begin{tabular}{llll}
% Model (abbrev.) & Model architecture               & Input         & Featurization scheme                  \\
% \hline
% Random forest (RF)  & Tree-based ensemble method & Vector & Voronoi tessellation + elemental attributes  \\
% XGBoost (XGB)       & Tree-based ensemble method & Vector & Voronoi tessellation + elemental attributes  \\
% GMP+single NN (GMP) & Regular neural networks    & Vector & Electron density from pseudopotentials \\
% ALIGNN (ALIGNN) & Graph neural networks            & Crystal graph & One-hot encoding elemental attributes \\
% LLM-Prop (LLM)  & Transformer-based language model & Text          & Robocrystallographer text generation 
% \end{tabular}
% \end{ruledtabular}
% \end{table*}

\subsection{Datasets and models}

We use the snapshots of the JARVIS, Materials Project, and OMQD databases used in our previous study~\cite{li2023exploiting}, available on Zenodo at \url{https://zenodo.org/records/8200972}. The snapshots correspond the JARVIS 2022.12.12 version, the Materials Project 2021.11.10 version, and the OQMD v1.6 version (published in November 2023), and they have been preprocessed to remove materials with formation energies larger than 5 eV/atom.

Performing hyperparameter tuning for each of over 700 OOD tasks for every model would be computationally impractical. In addition, hyperparameter tuning techniques are developed mainly to optimize the in-distribution rather than OOD performance. Our strategy is to perform the hyperparameter tuning on an in-distribution validation set randomly sampled from the whole dataset; then we apply the same optimal hyperparameters to every OOD task. Here, the criterion for the optimal hyperparameters is based on not only model performance but also training cost, to make the training amenable to extensive benchmarking. Our objective is to demonstrate the overall generalization performance on various OOD tasks rather than ensure optimal performance on each OOD task for every model. 
% For instance, we expect that model performance on the in-domain OOD tasks may be further improved by using optimal hyperparameters tuned on in-distribution data. 
We expect future work to develop techniques to optimize OOD performance by focusing on a small set of representative tasks selected based on this study.

% Todo: add a slurb for models?

For tree-based models, we use a descriptor set that consists of 145 compositional features~\cite{Ward2016} extracted from chemical formula and 128 structural features~\cite{ward2017including} extracted from crystal structures. For the RF model~\cite{sklearn}, we use 100 estimators, 30~\% of the features for the best splitting, and default settings for other hyperparameters. For the XGB model~\cite{xgboost}, we use 4 parallel boosting trees; for each boosting tree, we use 1000 estimators, a learning rate of 0.25, an L1 (L2) regularization strength of 0.01 (0.1), and the histogram tree grow method; we set the subsample ratio of training instances to 0.85, the subsample ratio of columns to 0.3 when constructing each tree, and the subsample ratio of columns to 0.5 for each level. 

The deep learning models (ALIGNN, GMP, and LLM-Prop) are trained as follows. ALIGNN models are trained with 2 ALIGNN layers, 2 GCN layers, a batch size of 16, 25 epochs, and layer normalization, while keeping other hyperparameters the same as in the original implementation~\cite{choudhary2021atomistic}. LLM-Prop models are trained with a batch size of 64, a max token length of 1500, and a drop rate of 0.5, while keeping the tokenizer, preprocessing strategy, and other hyperparameters as in the original implementation~\cite{rubungo2023llm}. For GMP models, we use the GMP-featurizer package~\cite{lei2023gmp} to calculate the Gaussian Multipole features by using 40 radial probes, max 4$^{th}$-order angular probes, and a Gaussian width of 1.5. The Kresse-Joubert projector augmented-wave pseudopotentials are used to calculated electron density~\cite{kresse1999ultrasoft}. The model architecture consists of a compressor neural network (hidden layer dimension [512,256,128,64,32]), a Set2Set neural network with 5 hidden layers to convert sets of varying sizes to a fixed size tensor~\cite{vinyals2015order}, and a predictor neural network (hidden layer dimension [256,128,64,32,16]). GMP models are trained with 200 epochs, a batch size of 128, and a ReduceLROnPlateau scheduler to tune the learning rate on a validation set (10~\% of the initial training set).

\subsection{SHAP analysis for leave-one-element-out tasks}

The following procedure is proposed to identify the sources of biases in the models trained for the leave-one-element-out tasks. For each task, we use the model trained without element $X$ to make predictions on the whole dataset. Then, we train a second model on the prediction errors of the first model on the whole dataset. By doing so, the second model learns to correct the bias in the first model which is the one used in the leave-one-element-out task. Finally, we calculate the SHAP  feature contributions of the second trained model for all data with element $X$, interpreted as the correction for the feature contribution of the first model for the out-of-distribution materials. Here the first and second models are chosen to be XGB because it uses two types of interpretable features: the compositional features are chemical attributes computed from chemical formula~\cite{Ward2016}, whereas the structural features are characteristics of the local atomic environment calculated from crystal structures~\cite{ward2017including}. We apply the above procedure to calculate the SHAP contributions for compositional and structural features of the second XGB model and aggregate the contributions into two categories. 

\subsection{UMAP dimension reduction}

The 2-component UMAP representation~\cite{umap} are built using a local neighborhood size of 50 and a minimum distance of 0.1. For deep learning models, the embeddings are used as inputs to construct the UMAP plots. For tree-based models, we sequentially drop highly correlated features with a Pearson correlation threshold of 0.7 and then construct the UMAP plots based on the final feature set. We use the kernel-density estimation method implemented in SciPy~\cite{virtanen2020scipy} to calculate the probability density of training data for every test data point. A probability density threshold of 0.001 is used to separate between in-domain and out-of-domain test data.

\section*{Data availability}
We use the same snapshots as in our prior work~\cite{li2023exploiting} for the JARVIS, MP, and OQMD datasets, which can be found on Zenodo at \url{https://zenodo.org/records/8200972}.

\section*{Code availability}
The code used for ML training and analysis is available on GitHub at (to be inserted upon acceptance).

\bibliography{lib}

\begin{thebibliography}{50}%
\makeatletter
\providecommand \@ifxundefined [1]{%
 \@ifx{#1\undefined}
}%
\providecommand \@ifnum [1]{%
 \ifnum #1\expandafter \@firstoftwo
 \else \expandafter \@secondoftwo
 \fi
}%
\providecommand \@ifx [1]{%
 \ifx #1\expandafter \@firstoftwo
 \else \expandafter \@secondoftwo
 \fi
}%
\providecommand \natexlab [1]{#1}%
\providecommand \enquote  [1]{``#1''}%
\providecommand \bibnamefont  [1]{#1}%
\providecommand \bibfnamefont [1]{#1}%
\providecommand \citenamefont [1]{#1}%
\providecommand \href@noop [0]{\@secondoftwo}%
\providecommand \href [0]{\begingroup \@sanitize@url \@href}%
\providecommand \@href[1]{\@@startlink{#1}\@@href}%
\providecommand \@@href[1]{\endgroup#1\@@endlink}%
\providecommand \@sanitize@url [0]{\catcode `\\12\catcode `\$12\catcode
  `\&12\catcode `\#12\catcode `\^12\catcode `\_12\catcode `\%12\relax}%
\providecommand \@@startlink[1]{}%
\providecommand \@@endlink[0]{}%
\providecommand \url  [0]{\begingroup\@sanitize@url \@url }%
\providecommand \@url [1]{\endgroup\@href {#1}{\urlprefix }}%
\providecommand \urlprefix  [0]{URL }%
\providecommand \Eprint [0]{\href }%
\providecommand \doibase [0]{https://doi.org/}%
\providecommand \selectlanguage [0]{\@gobble}%
\providecommand \bibinfo  [0]{\@secondoftwo}%
\providecommand \bibfield  [0]{\@secondoftwo}%
\providecommand \translation [1]{[#1]}%
\providecommand \BibitemOpen [0]{}%
\providecommand \bibitemStop [0]{}%
\providecommand \bibitemNoStop [0]{.\EOS\space}%
\providecommand \EOS [0]{\spacefactor3000\relax}%
\providecommand \BibitemShut  [1]{\csname bibitem#1\endcsname}%
\let\auto@bib@innerbib\@empty
%</preamble>
\bibitem [{\citenamefont {Butler}\ \emph {et~al.}(2018)\citenamefont {Butler},
  \citenamefont {Davies}, \citenamefont {Cartwright}, \citenamefont {Isayev},\
  and\ \citenamefont {Walsh}}]{butler2018machine}%
  \BibitemOpen
  \bibfield  {author} {\bibinfo {author} {\bibfnamefont {K.~T.}\ \bibnamefont
  {Butler}}, \bibinfo {author} {\bibfnamefont {D.~W.}\ \bibnamefont {Davies}},
  \bibinfo {author} {\bibfnamefont {H.}~\bibnamefont {Cartwright}}, \bibinfo
  {author} {\bibfnamefont {O.}~\bibnamefont {Isayev}},\ and\ \bibinfo {author}
  {\bibfnamefont {A.}~\bibnamefont {Walsh}},\ }\bibfield  {title} {\bibinfo
  {title} {Machine learning for molecular and materials science},\ }\href@noop
  {} {\bibfield  {journal} {\bibinfo  {journal} {Nature}\ }\textbf {\bibinfo
  {volume} {559}},\ \bibinfo {pages} {547} (\bibinfo {year}
  {2018})}\BibitemShut {NoStop}%
\bibitem [{\citenamefont {Carleo}\ \emph {et~al.}(2019)\citenamefont {Carleo},
  \citenamefont {Cirac}, \citenamefont {Cranmer}, \citenamefont {Daudet},
  \citenamefont {Schuld}, \citenamefont {Tishby}, \citenamefont
  {Vogt-Maranto},\ and\ \citenamefont {Zdeborov{\'a}}}]{carleo2019machine}%
  \BibitemOpen
  \bibfield  {author} {\bibinfo {author} {\bibfnamefont {G.}~\bibnamefont
  {Carleo}}, \bibinfo {author} {\bibfnamefont {I.}~\bibnamefont {Cirac}},
  \bibinfo {author} {\bibfnamefont {K.}~\bibnamefont {Cranmer}}, \bibinfo
  {author} {\bibfnamefont {L.}~\bibnamefont {Daudet}}, \bibinfo {author}
  {\bibfnamefont {M.}~\bibnamefont {Schuld}}, \bibinfo {author} {\bibfnamefont
  {N.}~\bibnamefont {Tishby}}, \bibinfo {author} {\bibfnamefont
  {L.}~\bibnamefont {Vogt-Maranto}},\ and\ \bibinfo {author} {\bibfnamefont
  {L.}~\bibnamefont {Zdeborov{\'a}}},\ }\bibfield  {title} {\bibinfo {title}
  {Machine learning and the physical sciences},\ }\href@noop {} {\bibfield
  {journal} {\bibinfo  {journal} {Reviews of Modern Physics}\ }\textbf
  {\bibinfo {volume} {91}},\ \bibinfo {pages} {045002} (\bibinfo {year}
  {2019})}\BibitemShut {NoStop}%
\bibitem [{\citenamefont {Krenn}\ \emph {et~al.}(2022)\citenamefont {Krenn},
  \citenamefont {Pollice}, \citenamefont {Guo}, \citenamefont {Aldeghi},
  \citenamefont {Cervera-Lierta}, \citenamefont {Friederich}, \citenamefont
  {dos Passos~Gomes}, \citenamefont {H{\"a}se}, \citenamefont {Jinich},
  \citenamefont {Nigam} \emph {et~al.}}]{krenn2022scientific}%
  \BibitemOpen
  \bibfield  {author} {\bibinfo {author} {\bibfnamefont {M.}~\bibnamefont
  {Krenn}}, \bibinfo {author} {\bibfnamefont {R.}~\bibnamefont {Pollice}},
  \bibinfo {author} {\bibfnamefont {S.~Y.}\ \bibnamefont {Guo}}, \bibinfo
  {author} {\bibfnamefont {M.}~\bibnamefont {Aldeghi}}, \bibinfo {author}
  {\bibfnamefont {A.}~\bibnamefont {Cervera-Lierta}}, \bibinfo {author}
  {\bibfnamefont {P.}~\bibnamefont {Friederich}}, \bibinfo {author}
  {\bibfnamefont {G.}~\bibnamefont {dos Passos~Gomes}}, \bibinfo {author}
  {\bibfnamefont {F.}~\bibnamefont {H{\"a}se}}, \bibinfo {author}
  {\bibfnamefont {A.}~\bibnamefont {Jinich}}, \bibinfo {author} {\bibfnamefont
  {A.}~\bibnamefont {Nigam}}, \emph {et~al.},\ }\bibfield  {title} {\bibinfo
  {title} {On scientific understanding with artificial intelligence},\
  }\href@noop {} {\bibfield  {journal} {\bibinfo  {journal} {Nature Reviews
  Physics}\ ,\ \bibinfo {pages} {1}} (\bibinfo {year} {2022})}\BibitemShut
  {NoStop}%
\bibitem [{\citenamefont {Huang}\ \emph {et~al.}(2023)\citenamefont {Huang},
  \citenamefont {von Rudorff},\ and\ \citenamefont {von
  Lilienfeld}}]{huang2023central}%
  \BibitemOpen
  \bibfield  {author} {\bibinfo {author} {\bibfnamefont {B.}~\bibnamefont
  {Huang}}, \bibinfo {author} {\bibfnamefont {G.~F.}\ \bibnamefont {von
  Rudorff}},\ and\ \bibinfo {author} {\bibfnamefont {O.~A.}\ \bibnamefont {von
  Lilienfeld}},\ }\bibfield  {title} {\bibinfo {title} {The central role of
  density functional theory in the ai age},\ }\href@noop {} {\bibfield
  {journal} {\bibinfo  {journal} {Science}\ }\textbf {\bibinfo {volume}
  {381}},\ \bibinfo {pages} {170} (\bibinfo {year} {2023})}\BibitemShut
  {NoStop}%
\bibitem [{\citenamefont {Liu}\ \emph {et~al.}(2021)\citenamefont {Liu},
  \citenamefont {Shen}, \citenamefont {He}, \citenamefont {Zhang},
  \citenamefont {Xu}, \citenamefont {Yu},\ and\ \citenamefont
  {Cui}}]{liu2021towards}%
  \BibitemOpen
  \bibfield  {author} {\bibinfo {author} {\bibfnamefont {J.}~\bibnamefont
  {Liu}}, \bibinfo {author} {\bibfnamefont {Z.}~\bibnamefont {Shen}}, \bibinfo
  {author} {\bibfnamefont {Y.}~\bibnamefont {He}}, \bibinfo {author}
  {\bibfnamefont {X.}~\bibnamefont {Zhang}}, \bibinfo {author} {\bibfnamefont
  {R.}~\bibnamefont {Xu}}, \bibinfo {author} {\bibfnamefont {H.}~\bibnamefont
  {Yu}},\ and\ \bibinfo {author} {\bibfnamefont {P.}~\bibnamefont {Cui}},\
  }\bibfield  {title} {\bibinfo {title} {Towards out-of-distribution
  generalization: A survey},\ }\href@noop {} {\bibfield  {journal} {\bibinfo
  {journal} {arXiv preprint arXiv:2108.13624}\ } (\bibinfo {year}
  {2021})}\BibitemShut {NoStop}%
\bibitem [{\citenamefont {Merchant}\ \emph {et~al.}(2023)\citenamefont
  {Merchant}, \citenamefont {Batzner}, \citenamefont {Schoenholz},
  \citenamefont {Aykol}, \citenamefont {Cheon},\ and\ \citenamefont
  {Cubuk}}]{merchant2023scaling}%
  \BibitemOpen
  \bibfield  {author} {\bibinfo {author} {\bibfnamefont {A.}~\bibnamefont
  {Merchant}}, \bibinfo {author} {\bibfnamefont {S.}~\bibnamefont {Batzner}},
  \bibinfo {author} {\bibfnamefont {S.~S.}\ \bibnamefont {Schoenholz}},
  \bibinfo {author} {\bibfnamefont {M.}~\bibnamefont {Aykol}}, \bibinfo
  {author} {\bibfnamefont {G.}~\bibnamefont {Cheon}},\ and\ \bibinfo {author}
  {\bibfnamefont {E.~D.}\ \bibnamefont {Cubuk}},\ }\bibfield  {title} {\bibinfo
  {title} {Scaling deep learning for materials discovery},\ }\href@noop {}
  {\bibfield  {journal} {\bibinfo  {journal} {Nature}\ ,\ \bibinfo {pages} {1}}
  (\bibinfo {year} {2023})}\BibitemShut {NoStop}%
\bibitem [{\citenamefont {Yang}\ \emph {et~al.}(2024)\citenamefont {Yang},
  \citenamefont {Hu}, \citenamefont {Zhou}, \citenamefont {Liu}, \citenamefont
  {Shi}, \citenamefont {Li}, \citenamefont {Li}, \citenamefont {Chen},
  \citenamefont {Chen}, \citenamefont {Zeni} \emph
  {et~al.}}]{yang2024mattersim}%
  \BibitemOpen
  \bibfield  {author} {\bibinfo {author} {\bibfnamefont {H.}~\bibnamefont
  {Yang}}, \bibinfo {author} {\bibfnamefont {C.}~\bibnamefont {Hu}}, \bibinfo
  {author} {\bibfnamefont {Y.}~\bibnamefont {Zhou}}, \bibinfo {author}
  {\bibfnamefont {X.}~\bibnamefont {Liu}}, \bibinfo {author} {\bibfnamefont
  {Y.}~\bibnamefont {Shi}}, \bibinfo {author} {\bibfnamefont {J.}~\bibnamefont
  {Li}}, \bibinfo {author} {\bibfnamefont {G.}~\bibnamefont {Li}}, \bibinfo
  {author} {\bibfnamefont {Z.}~\bibnamefont {Chen}}, \bibinfo {author}
  {\bibfnamefont {S.}~\bibnamefont {Chen}}, \bibinfo {author} {\bibfnamefont
  {C.}~\bibnamefont {Zeni}}, \emph {et~al.},\ }\bibfield  {title} {\bibinfo
  {title} {Mattersim: A deep learning atomistic model across elements,
  temperatures and pressures},\ }\href@noop {} {\bibfield  {journal} {\bibinfo
  {journal} {arXiv preprint arXiv:2405.04967}\ } (\bibinfo {year}
  {2024})}\BibitemShut {NoStop}%
\bibitem [{\citenamefont {Batatia}\ \emph {et~al.}(2023)\citenamefont
  {Batatia}, \citenamefont {Benner}, \citenamefont {Chiang}, \citenamefont
  {Elena}, \citenamefont {Kov{\'a}cs}, \citenamefont {Riebesell}, \citenamefont
  {Advincula}, \citenamefont {Asta}, \citenamefont {Baldwin}, \citenamefont
  {Bernstein} \emph {et~al.}}]{batatia2023foundation}%
  \BibitemOpen
  \bibfield  {author} {\bibinfo {author} {\bibfnamefont {I.}~\bibnamefont
  {Batatia}}, \bibinfo {author} {\bibfnamefont {P.}~\bibnamefont {Benner}},
  \bibinfo {author} {\bibfnamefont {Y.}~\bibnamefont {Chiang}}, \bibinfo
  {author} {\bibfnamefont {A.~M.}\ \bibnamefont {Elena}}, \bibinfo {author}
  {\bibfnamefont {D.~P.}\ \bibnamefont {Kov{\'a}cs}}, \bibinfo {author}
  {\bibfnamefont {J.}~\bibnamefont {Riebesell}}, \bibinfo {author}
  {\bibfnamefont {X.~R.}\ \bibnamefont {Advincula}}, \bibinfo {author}
  {\bibfnamefont {M.}~\bibnamefont {Asta}}, \bibinfo {author} {\bibfnamefont
  {W.~J.}\ \bibnamefont {Baldwin}}, \bibinfo {author} {\bibfnamefont
  {N.}~\bibnamefont {Bernstein}}, \emph {et~al.},\ }\bibfield  {title}
  {\bibinfo {title} {A foundation model for atomistic materials chemistry},\
  }\href@noop {} {\bibfield  {journal} {\bibinfo  {journal} {arXiv preprint
  arXiv:2401.00096}\ } (\bibinfo {year} {2023})}\BibitemShut {NoStop}%
\bibitem [{\citenamefont {Schmidt}\ \emph {et~al.}(2022)\citenamefont
  {Schmidt}, \citenamefont {Hoffmann}, \citenamefont {Wang}, \citenamefont
  {Borlido}, \citenamefont {Carri{\c{c}}o}, \citenamefont {Cerqueira},
  \citenamefont {Botti},\ and\ \citenamefont {Marques}}]{schmidt2022large}%
  \BibitemOpen
  \bibfield  {author} {\bibinfo {author} {\bibfnamefont {J.}~\bibnamefont
  {Schmidt}}, \bibinfo {author} {\bibfnamefont {N.}~\bibnamefont {Hoffmann}},
  \bibinfo {author} {\bibfnamefont {H.-C.}\ \bibnamefont {Wang}}, \bibinfo
  {author} {\bibfnamefont {P.}~\bibnamefont {Borlido}}, \bibinfo {author}
  {\bibfnamefont {P.~J.}\ \bibnamefont {Carri{\c{c}}o}}, \bibinfo {author}
  {\bibfnamefont {T.~F.}\ \bibnamefont {Cerqueira}}, \bibinfo {author}
  {\bibfnamefont {S.}~\bibnamefont {Botti}},\ and\ \bibinfo {author}
  {\bibfnamefont {M.~A.}\ \bibnamefont {Marques}},\ }\bibfield  {title}
  {\bibinfo {title} {Large-scale machine-learning-assisted exploration of the
  whole materials space},\ }\href@noop {} {\bibfield  {journal} {\bibinfo
  {journal} {arXiv preprint arXiv:2210.00579}\ } (\bibinfo {year}
  {2022})}\BibitemShut {NoStop}%
\bibitem [{\citenamefont {Deng}\ \emph {et~al.}(2023)\citenamefont {Deng},
  \citenamefont {Zhong}, \citenamefont {Jun}, \citenamefont {Riebesell},
  \citenamefont {Han}, \citenamefont {Bartel},\ and\ \citenamefont
  {Ceder}}]{deng2023chgnet}%
  \BibitemOpen
  \bibfield  {author} {\bibinfo {author} {\bibfnamefont {B.}~\bibnamefont
  {Deng}}, \bibinfo {author} {\bibfnamefont {P.}~\bibnamefont {Zhong}},
  \bibinfo {author} {\bibfnamefont {K.}~\bibnamefont {Jun}}, \bibinfo {author}
  {\bibfnamefont {J.}~\bibnamefont {Riebesell}}, \bibinfo {author}
  {\bibfnamefont {K.}~\bibnamefont {Han}}, \bibinfo {author} {\bibfnamefont
  {C.~J.}\ \bibnamefont {Bartel}},\ and\ \bibinfo {author} {\bibfnamefont
  {G.}~\bibnamefont {Ceder}},\ }\bibfield  {title} {\bibinfo {title} {Chgnet as
  a pretrained universal neural network potential for charge-informed atomistic
  modelling},\ }\href@noop {} {\bibfield  {journal} {\bibinfo  {journal}
  {Nature Machine Intelligence}\ }\textbf {\bibinfo {volume} {5}},\ \bibinfo
  {pages} {1031} (\bibinfo {year} {2023})}\BibitemShut {NoStop}%
\bibitem [{\citenamefont {Chen}\ and\ \citenamefont
  {Ong}(2022)}]{chen2022universal}%
  \BibitemOpen
  \bibfield  {author} {\bibinfo {author} {\bibfnamefont {C.}~\bibnamefont
  {Chen}}\ and\ \bibinfo {author} {\bibfnamefont {S.~P.}\ \bibnamefont {Ong}},\
  }\bibfield  {title} {\bibinfo {title} {A universal graph deep learning
  interatomic potential for the periodic table},\ }\href@noop {} {\bibfield
  {journal} {\bibinfo  {journal} {Nature Computational Science}\ }\textbf
  {\bibinfo {volume} {2}},\ \bibinfo {pages} {718} (\bibinfo {year}
  {2022})}\BibitemShut {NoStop}%
\bibitem [{\citenamefont {Li}\ \emph {et~al.}(2024)\citenamefont {Li},
  \citenamefont {Choudhary}, \citenamefont {DeCost}, \citenamefont
  {Greenwood},\ and\ \citenamefont {Hattrick-Simpers}}]{li2024efficient}%
  \BibitemOpen
  \bibfield  {author} {\bibinfo {author} {\bibfnamefont {K.}~\bibnamefont
  {Li}}, \bibinfo {author} {\bibfnamefont {K.}~\bibnamefont {Choudhary}},
  \bibinfo {author} {\bibfnamefont {B.}~\bibnamefont {DeCost}}, \bibinfo
  {author} {\bibfnamefont {M.}~\bibnamefont {Greenwood}},\ and\ \bibinfo
  {author} {\bibfnamefont {J.}~\bibnamefont {Hattrick-Simpers}},\ }\bibfield
  {title} {\bibinfo {title} {Efficient first principles based modeling via
  machine learning: from simple representations to high entropy materials},\
  }\href@noop {} {\bibfield  {journal} {\bibinfo  {journal} {Journal of
  Materials Chemistry A}\ } (\bibinfo {year} {2024})}\BibitemShut {NoStop}%
\bibitem [{\citenamefont {Chen}\ \emph {et~al.}(2023)\citenamefont {Chen},
  \citenamefont {Hilhorst}, \citenamefont {Bokas}, \citenamefont {Gorsse},
  \citenamefont {Jacques},\ and\ \citenamefont {Hautier}}]{chen2023map}%
  \BibitemOpen
  \bibfield  {author} {\bibinfo {author} {\bibfnamefont {W.}~\bibnamefont
  {Chen}}, \bibinfo {author} {\bibfnamefont {A.}~\bibnamefont {Hilhorst}},
  \bibinfo {author} {\bibfnamefont {G.}~\bibnamefont {Bokas}}, \bibinfo
  {author} {\bibfnamefont {S.}~\bibnamefont {Gorsse}}, \bibinfo {author}
  {\bibfnamefont {P.~J.}\ \bibnamefont {Jacques}},\ and\ \bibinfo {author}
  {\bibfnamefont {G.}~\bibnamefont {Hautier}},\ }\bibfield  {title} {\bibinfo
  {title} {A map of single-phase high-entropy alloys},\ }\href@noop {}
  {\bibfield  {journal} {\bibinfo  {journal} {Nature Communications}\ }\textbf
  {\bibinfo {volume} {14}},\ \bibinfo {pages} {2856} (\bibinfo {year}
  {2023})}\BibitemShut {NoStop}%
\bibitem [{\citenamefont {Bokas}\ \emph {et~al.}(2021)\citenamefont {Bokas},
  \citenamefont {Chen}, \citenamefont {Hilhorst}, \citenamefont {Jacques},
  \citenamefont {Gorsse},\ and\ \citenamefont {Hautier}}]{bokas2021unveiling}%
  \BibitemOpen
  \bibfield  {author} {\bibinfo {author} {\bibfnamefont {G.~B.}\ \bibnamefont
  {Bokas}}, \bibinfo {author} {\bibfnamefont {W.}~\bibnamefont {Chen}},
  \bibinfo {author} {\bibfnamefont {A.}~\bibnamefont {Hilhorst}}, \bibinfo
  {author} {\bibfnamefont {P.~J.}\ \bibnamefont {Jacques}}, \bibinfo {author}
  {\bibfnamefont {S.}~\bibnamefont {Gorsse}},\ and\ \bibinfo {author}
  {\bibfnamefont {G.}~\bibnamefont {Hautier}},\ }\bibfield  {title} {\bibinfo
  {title} {Unveiling the thermodynamic driving forces for high entropy alloys
  formation through big data ab initio analysis},\ }\href@noop {} {\bibfield
  {journal} {\bibinfo  {journal} {Scripta Materialia}\ }\textbf {\bibinfo
  {volume} {202}},\ \bibinfo {pages} {114000} (\bibinfo {year}
  {2021})}\BibitemShut {NoStop}%
\bibitem [{\citenamefont {Chen}\ \emph {et~al.}(2018)\citenamefont {Chen},
  \citenamefont {Mao},\ and\ \citenamefont {Chen}}]{chen2018database}%
  \BibitemOpen
  \bibfield  {author} {\bibinfo {author} {\bibfnamefont {H.-L.}\ \bibnamefont
  {Chen}}, \bibinfo {author} {\bibfnamefont {H.}~\bibnamefont {Mao}},\ and\
  \bibinfo {author} {\bibfnamefont {Q.}~\bibnamefont {Chen}},\ }\bibfield
  {title} {\bibinfo {title} {Database development and calphad calculations for
  high entropy alloys: Challenges, strategies, and tips},\ }\href@noop {}
  {\bibfield  {journal} {\bibinfo  {journal} {Materials Chemistry and Physics}\
  }\textbf {\bibinfo {volume} {210}},\ \bibinfo {pages} {279} (\bibinfo {year}
  {2018})}\BibitemShut {NoStop}%
\bibitem [{\citenamefont {Frey}\ \emph {et~al.}(2023)\citenamefont {Frey},
  \citenamefont {Soklaski}, \citenamefont {Axelrod}, \citenamefont {Samsi},
  \citenamefont {Gomez-Bombarelli}, \citenamefont {Coley},\ and\ \citenamefont
  {Gadepally}}]{frey2023neural}%
  \BibitemOpen
  \bibfield  {author} {\bibinfo {author} {\bibfnamefont {N.~C.}\ \bibnamefont
  {Frey}}, \bibinfo {author} {\bibfnamefont {R.}~\bibnamefont {Soklaski}},
  \bibinfo {author} {\bibfnamefont {S.}~\bibnamefont {Axelrod}}, \bibinfo
  {author} {\bibfnamefont {S.}~\bibnamefont {Samsi}}, \bibinfo {author}
  {\bibfnamefont {R.}~\bibnamefont {Gomez-Bombarelli}}, \bibinfo {author}
  {\bibfnamefont {C.~W.}\ \bibnamefont {Coley}},\ and\ \bibinfo {author}
  {\bibfnamefont {V.}~\bibnamefont {Gadepally}},\ }\bibfield  {title} {\bibinfo
  {title} {Neural scaling of deep chemical models},\ }\href@noop {} {\bibfield
  {journal} {\bibinfo  {journal} {Nature Machine Intelligence}\ }\textbf
  {\bibinfo {volume} {5}},\ \bibinfo {pages} {1297} (\bibinfo {year}
  {2023})}\BibitemShut {NoStop}%
\bibitem [{\citenamefont {Kaplan}\ \emph {et~al.}(2020)\citenamefont {Kaplan},
  \citenamefont {McCandlish}, \citenamefont {Henighan}, \citenamefont {Brown},
  \citenamefont {Chess}, \citenamefont {Child}, \citenamefont {Gray},
  \citenamefont {Radford}, \citenamefont {Wu},\ and\ \citenamefont
  {Amodei}}]{kaplan2020scaling}%
  \BibitemOpen
  \bibfield  {author} {\bibinfo {author} {\bibfnamefont {J.}~\bibnamefont
  {Kaplan}}, \bibinfo {author} {\bibfnamefont {S.}~\bibnamefont {McCandlish}},
  \bibinfo {author} {\bibfnamefont {T.}~\bibnamefont {Henighan}}, \bibinfo
  {author} {\bibfnamefont {T.~B.}\ \bibnamefont {Brown}}, \bibinfo {author}
  {\bibfnamefont {B.}~\bibnamefont {Chess}}, \bibinfo {author} {\bibfnamefont
  {R.}~\bibnamefont {Child}}, \bibinfo {author} {\bibfnamefont
  {S.}~\bibnamefont {Gray}}, \bibinfo {author} {\bibfnamefont {A.}~\bibnamefont
  {Radford}}, \bibinfo {author} {\bibfnamefont {J.}~\bibnamefont {Wu}},\ and\
  \bibinfo {author} {\bibfnamefont {D.}~\bibnamefont {Amodei}},\ }\bibfield
  {title} {\bibinfo {title} {Scaling laws for neural language models},\
  }\href@noop {} {\bibfield  {journal} {\bibinfo  {journal} {arXiv preprint
  arXiv:2001.08361}\ } (\bibinfo {year} {2020})}\BibitemShut {NoStop}%
\bibitem [{\citenamefont {Choudhary}\ \emph {et~al.}(2020)\citenamefont
  {Choudhary}, \citenamefont {Garrity}, \citenamefont {Reid}, \citenamefont
  {DeCost}, \citenamefont {Biacchi}, \citenamefont {Hight~Walker},
  \citenamefont {Trautt}, \citenamefont {Hattrick-Simpers}, \citenamefont
  {Kusne}, \citenamefont {Centrone} \emph {et~al.}}]{choudhary2020joint}%
  \BibitemOpen
  \bibfield  {author} {\bibinfo {author} {\bibfnamefont {K.}~\bibnamefont
  {Choudhary}}, \bibinfo {author} {\bibfnamefont {K.~F.}\ \bibnamefont
  {Garrity}}, \bibinfo {author} {\bibfnamefont {A.~C.}\ \bibnamefont {Reid}},
  \bibinfo {author} {\bibfnamefont {B.}~\bibnamefont {DeCost}}, \bibinfo
  {author} {\bibfnamefont {A.~J.}\ \bibnamefont {Biacchi}}, \bibinfo {author}
  {\bibfnamefont {A.~R.}\ \bibnamefont {Hight~Walker}}, \bibinfo {author}
  {\bibfnamefont {Z.}~\bibnamefont {Trautt}}, \bibinfo {author} {\bibfnamefont
  {J.}~\bibnamefont {Hattrick-Simpers}}, \bibinfo {author} {\bibfnamefont
  {A.~G.}\ \bibnamefont {Kusne}}, \bibinfo {author} {\bibfnamefont
  {A.}~\bibnamefont {Centrone}}, \emph {et~al.},\ }\bibfield  {title} {\bibinfo
  {title} {The joint automated repository for various integrated simulations
  (jarvis) for data-driven materials design},\ }\href@noop {} {\bibfield
  {journal} {\bibinfo  {journal} {npj computational materials}\ }\textbf
  {\bibinfo {volume} {6}},\ \bibinfo {pages} {173} (\bibinfo {year}
  {2020})}\BibitemShut {NoStop}%
\bibitem [{\citenamefont {Wines}\ \emph {et~al.}(2023)\citenamefont {Wines},
  \citenamefont {Gurunathan}, \citenamefont {Garrity}, \citenamefont {DeCost},
  \citenamefont {Biacchi}, \citenamefont {Tavazza},\ and\ \citenamefont
  {Choudhary}}]{Wines2023}%
  \BibitemOpen
  \bibfield  {author} {\bibinfo {author} {\bibfnamefont {D.}~\bibnamefont
  {Wines}}, \bibinfo {author} {\bibfnamefont {R.}~\bibnamefont {Gurunathan}},
  \bibinfo {author} {\bibfnamefont {K.~F.}\ \bibnamefont {Garrity}}, \bibinfo
  {author} {\bibfnamefont {B.}~\bibnamefont {DeCost}}, \bibinfo {author}
  {\bibfnamefont {A.~J.}\ \bibnamefont {Biacchi}}, \bibinfo {author}
  {\bibfnamefont {F.}~\bibnamefont {Tavazza}},\ and\ \bibinfo {author}
  {\bibfnamefont {K.}~\bibnamefont {Choudhary}},\ }\bibfield  {title} {\bibinfo
  {title} {Recent progress in the jarvis infrastructure for next-generation
  data-driven materials design},\ }\bibfield  {journal} {\bibinfo  {journal}
  {Applied Physics Reviews}\ }\textbf {\bibinfo {volume} {10}},\ \href
  {https://doi.org/10.1063/5.0159299} {10.1063/5.0159299} (\bibinfo {year}
  {2023})\BibitemShut {NoStop}%
\bibitem [{\citenamefont {Jain}\ \emph {et~al.}(2013)\citenamefont {Jain},
  \citenamefont {Ong}, \citenamefont {Hautier}, \citenamefont {Chen},
  \citenamefont {Richards}, \citenamefont {Dacek}, \citenamefont {Cholia},
  \citenamefont {Gunter}, \citenamefont {Skinner}, \citenamefont {Ceder} \emph
  {et~al.}}]{jain2013commentary}%
  \BibitemOpen
  \bibfield  {author} {\bibinfo {author} {\bibfnamefont {A.}~\bibnamefont
  {Jain}}, \bibinfo {author} {\bibfnamefont {S.~P.}\ \bibnamefont {Ong}},
  \bibinfo {author} {\bibfnamefont {G.}~\bibnamefont {Hautier}}, \bibinfo
  {author} {\bibfnamefont {W.}~\bibnamefont {Chen}}, \bibinfo {author}
  {\bibfnamefont {W.~D.}\ \bibnamefont {Richards}}, \bibinfo {author}
  {\bibfnamefont {S.}~\bibnamefont {Dacek}}, \bibinfo {author} {\bibfnamefont
  {S.}~\bibnamefont {Cholia}}, \bibinfo {author} {\bibfnamefont
  {D.}~\bibnamefont {Gunter}}, \bibinfo {author} {\bibfnamefont
  {D.}~\bibnamefont {Skinner}}, \bibinfo {author} {\bibfnamefont
  {G.}~\bibnamefont {Ceder}}, \emph {et~al.},\ }\bibfield  {title} {\bibinfo
  {title} {Commentary: The materials project: A materials genome approach to
  accelerating materials innovation},\ }\href@noop {} {\bibfield  {journal}
  {\bibinfo  {journal} {APL materials}\ }\textbf {\bibinfo {volume} {1}}
  (\bibinfo {year} {2013})}\BibitemShut {NoStop}%
\bibitem [{\citenamefont {Saal}\ \emph {et~al.}(2013)\citenamefont {Saal},
  \citenamefont {Kirklin}, \citenamefont {Aykol}, \citenamefont {Meredig},\
  and\ \citenamefont {Wolverton}}]{Saal2013}%
  \BibitemOpen
  \bibfield  {author} {\bibinfo {author} {\bibfnamefont {J.~E.}\ \bibnamefont
  {Saal}}, \bibinfo {author} {\bibfnamefont {S.}~\bibnamefont {Kirklin}},
  \bibinfo {author} {\bibfnamefont {M.}~\bibnamefont {Aykol}}, \bibinfo
  {author} {\bibfnamefont {B.}~\bibnamefont {Meredig}},\ and\ \bibinfo {author}
  {\bibfnamefont {C.}~\bibnamefont {Wolverton}},\ }\bibfield  {title} {\bibinfo
  {title} {{Materials design and discovery with high-throughput density
  functional theory: The open quantum materials database (OQMD)}},\ }\href
  {https://doi.org/10.1007/s11837-013-0755-4} {\bibfield  {journal} {\bibinfo
  {journal} {Jom}\ }\textbf {\bibinfo {volume} {65}},\ \bibinfo {pages} {1501}
  (\bibinfo {year} {2013})}\BibitemShut {NoStop}%
\bibitem [{\citenamefont {Breiman}(2001)}]{breiman2001random}%
  \BibitemOpen
  \bibfield  {author} {\bibinfo {author} {\bibfnamefont {L.}~\bibnamefont
  {Breiman}},\ }\bibfield  {title} {\bibinfo {title} {Random forests},\
  }\href@noop {} {\bibfield  {journal} {\bibinfo  {journal} {Machine learning}\
  }\textbf {\bibinfo {volume} {45}},\ \bibinfo {pages} {5} (\bibinfo {year}
  {2001})}\BibitemShut {NoStop}%
\bibitem [{\citenamefont {Chen}\ and\ \citenamefont
  {Guestrin}(2016)}]{xgboost}%
  \BibitemOpen
  \bibfield  {author} {\bibinfo {author} {\bibfnamefont {T.}~\bibnamefont
  {Chen}}\ and\ \bibinfo {author} {\bibfnamefont {C.}~\bibnamefont
  {Guestrin}},\ }\bibfield  {title} {\bibinfo {title} {{XGBoost}},\ }in\ \href
  {https://doi.org/10.1145/2939672.2939785} {\emph {\bibinfo {booktitle} {Proc.
  22nd ACM SIGKDD Int. Conf. Knowl. Discov. Data Min.}}}\ (\bibinfo
  {publisher} {ACM},\ \bibinfo {address} {New York, NY, USA},\ \bibinfo {year}
  {2016})\ pp.\ \bibinfo {pages} {785--794}\BibitemShut {NoStop}%
\bibitem [{\citenamefont {Ward}\ \emph {et~al.}(2018)\citenamefont {Ward},
  \citenamefont {Dunn}, \citenamefont {Faghaninia}, \citenamefont {Zimmermann},
  \citenamefont {Bajaj}, \citenamefont {Wang}, \citenamefont {Montoya},
  \citenamefont {Chen}, \citenamefont {Bystrom}, \citenamefont {Dylla} \emph
  {et~al.}}]{ward2018matminer}%
  \BibitemOpen
  \bibfield  {author} {\bibinfo {author} {\bibfnamefont {L.}~\bibnamefont
  {Ward}}, \bibinfo {author} {\bibfnamefont {A.}~\bibnamefont {Dunn}}, \bibinfo
  {author} {\bibfnamefont {A.}~\bibnamefont {Faghaninia}}, \bibinfo {author}
  {\bibfnamefont {N.~E.}\ \bibnamefont {Zimmermann}}, \bibinfo {author}
  {\bibfnamefont {S.}~\bibnamefont {Bajaj}}, \bibinfo {author} {\bibfnamefont
  {Q.}~\bibnamefont {Wang}}, \bibinfo {author} {\bibfnamefont {J.}~\bibnamefont
  {Montoya}}, \bibinfo {author} {\bibfnamefont {J.}~\bibnamefont {Chen}},
  \bibinfo {author} {\bibfnamefont {K.}~\bibnamefont {Bystrom}}, \bibinfo
  {author} {\bibfnamefont {M.}~\bibnamefont {Dylla}}, \emph {et~al.},\
  }\bibfield  {title} {\bibinfo {title} {Matminer: An open source toolkit for
  materials data mining},\ }\href@noop {} {\bibfield  {journal} {\bibinfo
  {journal} {Computational Materials Science}\ }\textbf {\bibinfo {volume}
  {152}},\ \bibinfo {pages} {60} (\bibinfo {year} {2018})}\BibitemShut
  {NoStop}%
\bibitem [{\citenamefont {Lei}\ and\ \citenamefont
  {Medford}(2022)}]{lei2022universal}%
  \BibitemOpen
  \bibfield  {author} {\bibinfo {author} {\bibfnamefont {X.}~\bibnamefont
  {Lei}}\ and\ \bibinfo {author} {\bibfnamefont {A.~J.}\ \bibnamefont
  {Medford}},\ }\bibfield  {title} {\bibinfo {title} {A universal framework for
  featurization of atomistic systems},\ }\href@noop {} {\bibfield  {journal}
  {\bibinfo  {journal} {The Journal of Physical Chemistry Letters}\ }\textbf
  {\bibinfo {volume} {13}},\ \bibinfo {pages} {7911} (\bibinfo {year}
  {2022})}\BibitemShut {NoStop}%
\bibitem [{\citenamefont {Choudhary}\ and\ \citenamefont
  {DeCost}(2021)}]{choudhary2021atomistic}%
  \BibitemOpen
  \bibfield  {author} {\bibinfo {author} {\bibfnamefont {K.}~\bibnamefont
  {Choudhary}}\ and\ \bibinfo {author} {\bibfnamefont {B.}~\bibnamefont
  {DeCost}},\ }\bibfield  {title} {\bibinfo {title} {Atomistic line graph
  neural network for improved materials property predictions},\ }\href@noop {}
  {\bibfield  {journal} {\bibinfo  {journal} {npj Computational Materials}\
  }\textbf {\bibinfo {volume} {7}},\ \bibinfo {pages} {185} (\bibinfo {year}
  {2021})}\BibitemShut {NoStop}%
\bibitem [{\citenamefont {Rubungo}\ \emph {et~al.}(2023)\citenamefont
  {Rubungo}, \citenamefont {Arnold}, \citenamefont {Rand},\ and\ \citenamefont
  {Dieng}}]{rubungo2023llm}%
  \BibitemOpen
  \bibfield  {author} {\bibinfo {author} {\bibfnamefont {A.~N.}\ \bibnamefont
  {Rubungo}}, \bibinfo {author} {\bibfnamefont {C.}~\bibnamefont {Arnold}},
  \bibinfo {author} {\bibfnamefont {B.~P.}\ \bibnamefont {Rand}},\ and\
  \bibinfo {author} {\bibfnamefont {A.~B.}\ \bibnamefont {Dieng}},\ }\bibfield
  {title} {\bibinfo {title} {Llm-prop: Predicting physical and electronic
  properties of crystalline solids from their text descriptions},\ }\href@noop
  {} {\bibfield  {journal} {\bibinfo  {journal} {arXiv preprint
  arXiv:2310.14029}\ } (\bibinfo {year} {2023})}\BibitemShut {NoStop}%
\bibitem [{\citenamefont {Riebesell}\ \emph {et~al.}(2022)\citenamefont
  {Riebesell}, \citenamefont {Goodall},\ and\ \citenamefont
  {Baird}}]{riebesell_pymatviz_2022}%
  \BibitemOpen
  \bibfield  {author} {\bibinfo {author} {\bibfnamefont {J.}~\bibnamefont
  {Riebesell}}, \bibinfo {author} {\bibfnamefont {R.}~\bibnamefont {Goodall}},\
  and\ \bibinfo {author} {\bibfnamefont {S.~G.}\ \bibnamefont {Baird}},\ }\href
  {https://doi.org/10.5281/zenodo.7486816} {\bibinfo {title} {Pymatviz:
  visualization toolkit for materials informatics}} (\bibinfo {year} {2022}),\
  \bibinfo {note} {https://github.com/janosh/pymatviz}\BibitemShut {NoStop}%
\bibitem [{\citenamefont {Deng}\ \emph {et~al.}(2024)\citenamefont {Deng},
  \citenamefont {Choi}, \citenamefont {Zhong}, \citenamefont {Riebesell},
  \citenamefont {Anand}, \citenamefont {Li}, \citenamefont {Jun}, \citenamefont
  {Persson},\ and\ \citenamefont {Ceder}}]{deng2024overcoming}%
  \BibitemOpen
  \bibfield  {author} {\bibinfo {author} {\bibfnamefont {B.}~\bibnamefont
  {Deng}}, \bibinfo {author} {\bibfnamefont {Y.}~\bibnamefont {Choi}}, \bibinfo
  {author} {\bibfnamefont {P.}~\bibnamefont {Zhong}}, \bibinfo {author}
  {\bibfnamefont {J.}~\bibnamefont {Riebesell}}, \bibinfo {author}
  {\bibfnamefont {S.}~\bibnamefont {Anand}}, \bibinfo {author} {\bibfnamefont
  {Z.}~\bibnamefont {Li}}, \bibinfo {author} {\bibfnamefont {K.}~\bibnamefont
  {Jun}}, \bibinfo {author} {\bibfnamefont {K.~A.}\ \bibnamefont {Persson}},\
  and\ \bibinfo {author} {\bibfnamefont {G.}~\bibnamefont {Ceder}},\ }\bibfield
   {title} {\bibinfo {title} {Overcoming systematic softening in universal
  machine learning interatomic potentials by fine-tuning},\ }\href@noop {}
  {\bibfield  {journal} {\bibinfo  {journal} {arXiv preprint arXiv:2405.07105}\
  } (\bibinfo {year} {2024})}\BibitemShut {NoStop}%
\bibitem [{\citenamefont {Choudhary}\ and\ \citenamefont
  {Sumpter}(2023)}]{choudhary2023can}%
  \BibitemOpen
  \bibfield  {author} {\bibinfo {author} {\bibfnamefont {K.}~\bibnamefont
  {Choudhary}}\ and\ \bibinfo {author} {\bibfnamefont {B.~G.}\ \bibnamefont
  {Sumpter}},\ }\bibfield  {title} {\bibinfo {title} {Can a deep-learning model
  make fast predictions of vacancy formation in diverse materials?},\
  }\href@noop {} {\bibfield  {journal} {\bibinfo  {journal} {AIP Advances}\
  }\textbf {\bibinfo {volume} {13}} (\bibinfo {year} {2023})}\BibitemShut
  {NoStop}%
\bibitem [{\citenamefont {Lundberg}\ \emph {et~al.}(2020)\citenamefont
  {Lundberg}, \citenamefont {Erion}, \citenamefont {Chen}, \citenamefont
  {DeGrave}, \citenamefont {Prutkin}, \citenamefont {Nair}, \citenamefont
  {Katz}, \citenamefont {Himmelfarb}, \citenamefont {Bansal},\ and\
  \citenamefont {Lee}}]{lundberg2020local2global}%
  \BibitemOpen
  \bibfield  {author} {\bibinfo {author} {\bibfnamefont {S.~M.}\ \bibnamefont
  {Lundberg}}, \bibinfo {author} {\bibfnamefont {G.}~\bibnamefont {Erion}},
  \bibinfo {author} {\bibfnamefont {H.}~\bibnamefont {Chen}}, \bibinfo {author}
  {\bibfnamefont {A.}~\bibnamefont {DeGrave}}, \bibinfo {author} {\bibfnamefont
  {J.~M.}\ \bibnamefont {Prutkin}}, \bibinfo {author} {\bibfnamefont
  {B.}~\bibnamefont {Nair}}, \bibinfo {author} {\bibfnamefont {R.}~\bibnamefont
  {Katz}}, \bibinfo {author} {\bibfnamefont {J.}~\bibnamefont {Himmelfarb}},
  \bibinfo {author} {\bibfnamefont {N.}~\bibnamefont {Bansal}},\ and\ \bibinfo
  {author} {\bibfnamefont {S.-I.}\ \bibnamefont {Lee}},\ }\bibfield  {title}
  {\bibinfo {title} {From local explanations to global understanding with
  explainable ai for trees},\ }\href@noop {} {\bibfield  {journal} {\bibinfo
  {journal} {Nature Machine Intelligence}\ }\textbf {\bibinfo {volume} {2}},\
  \bibinfo {pages} {2522} (\bibinfo {year} {2020})}\BibitemShut {NoStop}%
\bibitem [{\citenamefont {Behler}\ and\ \citenamefont
  {Parrinello}(2007)}]{behler2007generalized}%
  \BibitemOpen
  \bibfield  {author} {\bibinfo {author} {\bibfnamefont {J.}~\bibnamefont
  {Behler}}\ and\ \bibinfo {author} {\bibfnamefont {M.}~\bibnamefont
  {Parrinello}},\ }\bibfield  {title} {\bibinfo {title} {Generalized
  neural-network representation of high-dimensional potential-energy
  surfaces},\ }\href@noop {} {\bibfield  {journal} {\bibinfo  {journal}
  {Physical review letters}\ }\textbf {\bibinfo {volume} {98}},\ \bibinfo
  {pages} {146401} (\bibinfo {year} {2007})}\BibitemShut {NoStop}%
\bibitem [{\citenamefont {Li}\ \emph {et~al.}(2023{\natexlab{a}})\citenamefont
  {Li}, \citenamefont {DeCost}, \citenamefont {Choudhary}, \citenamefont
  {Greenwood},\ and\ \citenamefont {Hattrick-Simpers}}]{li2022critical}%
  \BibitemOpen
  \bibfield  {author} {\bibinfo {author} {\bibfnamefont {K.}~\bibnamefont
  {Li}}, \bibinfo {author} {\bibfnamefont {B.}~\bibnamefont {DeCost}}, \bibinfo
  {author} {\bibfnamefont {K.}~\bibnamefont {Choudhary}}, \bibinfo {author}
  {\bibfnamefont {M.}~\bibnamefont {Greenwood}},\ and\ \bibinfo {author}
  {\bibfnamefont {J.}~\bibnamefont {Hattrick-Simpers}},\ }\bibfield  {title}
  {\bibinfo {title} {A critical examination of robustness and generalizability
  of machine learning prediction of materials properties},\ }\href@noop {}
  {\bibfield  {journal} {\bibinfo  {journal} {npj Computational Materials}\
  }\textbf {\bibinfo {volume} {9}},\ \bibinfo {pages} {55} (\bibinfo {year}
  {2023}{\natexlab{a}})}\BibitemShut {NoStop}%
\bibitem [{\citenamefont {Zhang}\ \emph {et~al.}(2023)\citenamefont {Zhang},
  \citenamefont {Chen}, \citenamefont {Rondinelli},\ and\ \citenamefont
  {Chen}}]{zhang2023entropy}%
  \BibitemOpen
  \bibfield  {author} {\bibinfo {author} {\bibfnamefont {H.}~\bibnamefont
  {Zhang}}, \bibinfo {author} {\bibfnamefont {W.~W.}\ \bibnamefont {Chen}},
  \bibinfo {author} {\bibfnamefont {J.~M.}\ \bibnamefont {Rondinelli}},\ and\
  \bibinfo {author} {\bibfnamefont {W.}~\bibnamefont {Chen}},\ }\bibfield
  {title} {\bibinfo {title} {Et-al: Entropy-targeted active learning for bias
  mitigation in materials data},\ }\href@noop {} {\bibfield  {journal}
  {\bibinfo  {journal} {Applied Physics Reviews}\ }\textbf {\bibinfo {volume}
  {10}} (\bibinfo {year} {2023})}\BibitemShut {NoStop}%
\bibitem [{\citenamefont {Schrier}\ \emph {et~al.}(2023)\citenamefont
  {Schrier}, \citenamefont {Norquist}, \citenamefont {Buonassisi},\ and\
  \citenamefont {Brgoch}}]{schrier2023pursuit}%
  \BibitemOpen
  \bibfield  {author} {\bibinfo {author} {\bibfnamefont {J.}~\bibnamefont
  {Schrier}}, \bibinfo {author} {\bibfnamefont {A.~J.}\ \bibnamefont
  {Norquist}}, \bibinfo {author} {\bibfnamefont {T.}~\bibnamefont
  {Buonassisi}},\ and\ \bibinfo {author} {\bibfnamefont {J.}~\bibnamefont
  {Brgoch}},\ }\bibfield  {title} {\bibinfo {title} {In pursuit of the
  exceptional: Research directions for machine learning in chemical and
  materials science},\ }\href@noop {} {\bibfield  {journal} {\bibinfo
  {journal} {Journal of the American Chemical Society}\ }\textbf {\bibinfo
  {volume} {145}},\ \bibinfo {pages} {21699} (\bibinfo {year}
  {2023})}\BibitemShut {NoStop}%
\bibitem [{\citenamefont {Goodall}\ \emph {et~al.}(2022)\citenamefont
  {Goodall}, \citenamefont {Parackal}, \citenamefont {Faber}, \citenamefont
  {Armiento},\ and\ \citenamefont {Lee}}]{goodall2022rapid}%
  \BibitemOpen
  \bibfield  {author} {\bibinfo {author} {\bibfnamefont {R.~E.}\ \bibnamefont
  {Goodall}}, \bibinfo {author} {\bibfnamefont {A.~S.}\ \bibnamefont
  {Parackal}}, \bibinfo {author} {\bibfnamefont {F.~A.}\ \bibnamefont {Faber}},
  \bibinfo {author} {\bibfnamefont {R.}~\bibnamefont {Armiento}},\ and\
  \bibinfo {author} {\bibfnamefont {A.~A.}\ \bibnamefont {Lee}},\ }\bibfield
  {title} {\bibinfo {title} {Rapid discovery of stable materials by
  coordinate-free coarse graining},\ }\href@noop {} {\bibfield  {journal}
  {\bibinfo  {journal} {Science Advances}\ }\textbf {\bibinfo {volume} {8}},\
  \bibinfo {pages} {eabn4117} (\bibinfo {year} {2022})}\BibitemShut {NoStop}%
\bibitem [{\citenamefont {Ganose}\ and\ \citenamefont
  {Jain}(2019)}]{ganose2019robocrystallographer}%
  \BibitemOpen
  \bibfield  {author} {\bibinfo {author} {\bibfnamefont {A.~M.}\ \bibnamefont
  {Ganose}}\ and\ \bibinfo {author} {\bibfnamefont {A.}~\bibnamefont {Jain}},\
  }\bibfield  {title} {\bibinfo {title} {Robocrystallographer: automated
  crystal structure text descriptions and analysis},\ }\href@noop {} {\bibfield
   {journal} {\bibinfo  {journal} {MRS Communications}\ }\textbf {\bibinfo
  {volume} {9}},\ \bibinfo {pages} {874} (\bibinfo {year} {2019})}\BibitemShut
  {NoStop}%
\bibitem [{\citenamefont {Li}\ \emph {et~al.}(2023{\natexlab{b}})\citenamefont
  {Li}, \citenamefont {Persaud}, \citenamefont {Choudhary}, \citenamefont
  {DeCost}, \citenamefont {Greenwood},\ and\ \citenamefont
  {Hattrick-Simpers}}]{li2023exploiting}%
  \BibitemOpen
  \bibfield  {author} {\bibinfo {author} {\bibfnamefont {K.}~\bibnamefont
  {Li}}, \bibinfo {author} {\bibfnamefont {D.}~\bibnamefont {Persaud}},
  \bibinfo {author} {\bibfnamefont {K.}~\bibnamefont {Choudhary}}, \bibinfo
  {author} {\bibfnamefont {B.}~\bibnamefont {DeCost}}, \bibinfo {author}
  {\bibfnamefont {M.}~\bibnamefont {Greenwood}},\ and\ \bibinfo {author}
  {\bibfnamefont {J.}~\bibnamefont {Hattrick-Simpers}},\ }\bibfield  {title}
  {\bibinfo {title} {Exploiting redundancy in large materials datasets for
  efficient machine learning with less data},\ }\href@noop {} {\bibfield
  {journal} {\bibinfo  {journal} {Nature Communications}\ }\textbf {\bibinfo
  {volume} {14}},\ \bibinfo {pages} {7283} (\bibinfo {year}
  {2023}{\natexlab{b}})}\BibitemShut {NoStop}%
\bibitem [{\citenamefont {McInnes}\ \emph {et~al.}(2018)\citenamefont
  {McInnes}, \citenamefont {Healy}, \citenamefont {Saul},\ and\ \citenamefont
  {Gro{\ss}berger}}]{umap}%
  \BibitemOpen
  \bibfield  {author} {\bibinfo {author} {\bibfnamefont {L.}~\bibnamefont
  {McInnes}}, \bibinfo {author} {\bibfnamefont {J.}~\bibnamefont {Healy}},
  \bibinfo {author} {\bibfnamefont {N.}~\bibnamefont {Saul}},\ and\ \bibinfo
  {author} {\bibfnamefont {L.}~\bibnamefont {Gro{\ss}berger}},\ }\bibfield
  {title} {\bibinfo {title} {{UMAP: Uniform Manifold Approximation and
  Projection}},\ }\href {https://doi.org/10.21105/joss.00861} {\bibfield
  {journal} {\bibinfo  {journal} {J. Open Source Softw.}\ }\textbf {\bibinfo
  {volume} {3}},\ \bibinfo {pages} {861} (\bibinfo {year} {2018})}\BibitemShut
  {NoStop}%
\bibitem [{\citenamefont {Riebesell}\ \emph {et~al.}(2023)\citenamefont
  {Riebesell}, \citenamefont {Goodall}, \citenamefont {Jain}, \citenamefont
  {Benner}, \citenamefont {Persson},\ and\ \citenamefont
  {Lee}}]{riebesell2023matbench}%
  \BibitemOpen
  \bibfield  {author} {\bibinfo {author} {\bibfnamefont {J.}~\bibnamefont
  {Riebesell}}, \bibinfo {author} {\bibfnamefont {R.~E.}\ \bibnamefont
  {Goodall}}, \bibinfo {author} {\bibfnamefont {A.}~\bibnamefont {Jain}},
  \bibinfo {author} {\bibfnamefont {P.}~\bibnamefont {Benner}}, \bibinfo
  {author} {\bibfnamefont {K.~A.}\ \bibnamefont {Persson}},\ and\ \bibinfo
  {author} {\bibfnamefont {A.~A.}\ \bibnamefont {Lee}},\ }\bibfield  {title}
  {\bibinfo {title} {Matbench discovery--an evaluation framework for machine
  learning crystal stability prediction},\ }\href@noop {} {\bibfield  {journal}
  {\bibinfo  {journal} {arXiv preprint arXiv:2308.14920}\ } (\bibinfo {year}
  {2023})}\BibitemShut {NoStop}%
\bibitem [{\citenamefont {Faber}\ \emph {et~al.}(2017)\citenamefont {Faber},
  \citenamefont {Hutchison}, \citenamefont {Huang}, \citenamefont {Gilmer},
  \citenamefont {Schoenholz}, \citenamefont {Dahl}, \citenamefont {Vinyals},
  \citenamefont {Kearnes}, \citenamefont {Riley},\ and\ \citenamefont
  {Von~Lilienfeld}}]{faber2017prediction}%
  \BibitemOpen
  \bibfield  {author} {\bibinfo {author} {\bibfnamefont {F.~A.}\ \bibnamefont
  {Faber}}, \bibinfo {author} {\bibfnamefont {L.}~\bibnamefont {Hutchison}},
  \bibinfo {author} {\bibfnamefont {B.}~\bibnamefont {Huang}}, \bibinfo
  {author} {\bibfnamefont {J.}~\bibnamefont {Gilmer}}, \bibinfo {author}
  {\bibfnamefont {S.~S.}\ \bibnamefont {Schoenholz}}, \bibinfo {author}
  {\bibfnamefont {G.~E.}\ \bibnamefont {Dahl}}, \bibinfo {author}
  {\bibfnamefont {O.}~\bibnamefont {Vinyals}}, \bibinfo {author} {\bibfnamefont
  {S.}~\bibnamefont {Kearnes}}, \bibinfo {author} {\bibfnamefont {P.~F.}\
  \bibnamefont {Riley}},\ and\ \bibinfo {author} {\bibfnamefont {O.~A.}\
  \bibnamefont {Von~Lilienfeld}},\ }\bibfield  {title} {\bibinfo {title}
  {Prediction errors of molecular machine learning models lower than hybrid dft
  error},\ }\href@noop {} {\bibfield  {journal} {\bibinfo  {journal} {Journal
  of chemical theory and computation}\ }\textbf {\bibinfo {volume} {13}},\
  \bibinfo {pages} {5255} (\bibinfo {year} {2017})}\BibitemShut {NoStop}%
\bibitem [{\citenamefont {Wei}\ \emph {et~al.}(2022)\citenamefont {Wei},
  \citenamefont {Tay}, \citenamefont {Bommasani}, \citenamefont {Raffel},
  \citenamefont {Zoph}, \citenamefont {Borgeaud}, \citenamefont {Yogatama},
  \citenamefont {Bosma}, \citenamefont {Zhou}, \citenamefont {Metzler} \emph
  {et~al.}}]{wei2022emergent}%
  \BibitemOpen
  \bibfield  {author} {\bibinfo {author} {\bibfnamefont {J.}~\bibnamefont
  {Wei}}, \bibinfo {author} {\bibfnamefont {Y.}~\bibnamefont {Tay}}, \bibinfo
  {author} {\bibfnamefont {R.}~\bibnamefont {Bommasani}}, \bibinfo {author}
  {\bibfnamefont {C.}~\bibnamefont {Raffel}}, \bibinfo {author} {\bibfnamefont
  {B.}~\bibnamefont {Zoph}}, \bibinfo {author} {\bibfnamefont {S.}~\bibnamefont
  {Borgeaud}}, \bibinfo {author} {\bibfnamefont {D.}~\bibnamefont {Yogatama}},
  \bibinfo {author} {\bibfnamefont {M.}~\bibnamefont {Bosma}}, \bibinfo
  {author} {\bibfnamefont {D.}~\bibnamefont {Zhou}}, \bibinfo {author}
  {\bibfnamefont {D.}~\bibnamefont {Metzler}}, \emph {et~al.},\ }\bibfield
  {title} {\bibinfo {title} {Emergent abilities of large language models},\
  }\href@noop {} {\bibfield  {journal} {\bibinfo  {journal} {arXiv preprint
  arXiv:2206.07682}\ } (\bibinfo {year} {2022})}\BibitemShut {NoStop}%
\bibitem [{\citenamefont {Biever}(2023)}]{biever2023chatgpt}%
  \BibitemOpen
  \bibfield  {author} {\bibinfo {author} {\bibfnamefont {C.}~\bibnamefont
  {Biever}},\ }\bibfield  {title} {\bibinfo {title} {Chatgpt broke the turing
  test-the race is on for new ways to assess ai},\ }\href@noop {} {\bibfield
  {journal} {\bibinfo  {journal} {Nature}\ }\textbf {\bibinfo {volume} {619}},\
  \bibinfo {pages} {686} (\bibinfo {year} {2023})}\BibitemShut {NoStop}%
\bibitem [{\citenamefont {Ward}\ \emph {et~al.}(2016)\citenamefont {Ward},
  \citenamefont {Agrawal}, \citenamefont {Choudhary},\ and\ \citenamefont
  {Wolverton}}]{Ward2016}%
  \BibitemOpen
  \bibfield  {author} {\bibinfo {author} {\bibfnamefont {L.}~\bibnamefont
  {Ward}}, \bibinfo {author} {\bibfnamefont {A.}~\bibnamefont {Agrawal}},
  \bibinfo {author} {\bibfnamefont {A.}~\bibnamefont {Choudhary}},\ and\
  \bibinfo {author} {\bibfnamefont {C.}~\bibnamefont {Wolverton}},\ }\bibfield
  {title} {\bibinfo {title} {{A general-purpose machine learning framework for
  predicting properties of inorganic materials}},\ }\href
  {https://doi.org/10.1038/npjcompumats.2016.28} {\bibfield  {journal}
  {\bibinfo  {journal} {npj Comput. Mater.}\ }\textbf {\bibinfo {volume} {2}},\
  \bibinfo {pages} {1} (\bibinfo {year} {2016})},\ \Eprint
  {https://arxiv.org/abs/1606.09551} {1606.09551} \BibitemShut {NoStop}%
\bibitem [{\citenamefont {Ward}\ \emph {et~al.}(2017)\citenamefont {Ward},
  \citenamefont {Liu}, \citenamefont {Krishna}, \citenamefont {Hegde},
  \citenamefont {Agrawal}, \citenamefont {Choudhary},\ and\ \citenamefont
  {Wolverton}}]{ward2017including}%
  \BibitemOpen
  \bibfield  {author} {\bibinfo {author} {\bibfnamefont {L.}~\bibnamefont
  {Ward}}, \bibinfo {author} {\bibfnamefont {R.}~\bibnamefont {Liu}}, \bibinfo
  {author} {\bibfnamefont {A.}~\bibnamefont {Krishna}}, \bibinfo {author}
  {\bibfnamefont {V.~I.}\ \bibnamefont {Hegde}}, \bibinfo {author}
  {\bibfnamefont {A.}~\bibnamefont {Agrawal}}, \bibinfo {author} {\bibfnamefont
  {A.}~\bibnamefont {Choudhary}},\ and\ \bibinfo {author} {\bibfnamefont
  {C.}~\bibnamefont {Wolverton}},\ }\bibfield  {title} {\bibinfo {title}
  {Including crystal structure attributes in machine learning models of
  formation energies via voronoi tessellations},\ }\href@noop {} {\bibfield
  {journal} {\bibinfo  {journal} {Physical Review B}\ }\textbf {\bibinfo
  {volume} {96}},\ \bibinfo {pages} {024104} (\bibinfo {year}
  {2017})}\BibitemShut {NoStop}%
\bibitem [{\citenamefont {Pedregosa}\ \emph {et~al.}(2011)\citenamefont
  {Pedregosa}, \citenamefont {Varoquaux}, \citenamefont {Gramfort},
  \citenamefont {Michel}, \citenamefont {Thirion}, \citenamefont {Grisel},
  \citenamefont {Blondel}, \citenamefont {Prettenhofer}, \citenamefont {Weiss},
  \citenamefont {Dubourg}, \citenamefont {Vanderplas}, \citenamefont {Passos},
  \citenamefont {Cournapeau}, \citenamefont {Brucher}, \citenamefont {Perrot},\
  and\ \citenamefont {Duchesnay}}]{sklearn}%
  \BibitemOpen
  \bibfield  {author} {\bibinfo {author} {\bibfnamefont {F.}~\bibnamefont
  {Pedregosa}}, \bibinfo {author} {\bibfnamefont {G.}~\bibnamefont
  {Varoquaux}}, \bibinfo {author} {\bibfnamefont {A.}~\bibnamefont {Gramfort}},
  \bibinfo {author} {\bibfnamefont {V.}~\bibnamefont {Michel}}, \bibinfo
  {author} {\bibfnamefont {B.}~\bibnamefont {Thirion}}, \bibinfo {author}
  {\bibfnamefont {O.}~\bibnamefont {Grisel}}, \bibinfo {author} {\bibfnamefont
  {M.}~\bibnamefont {Blondel}}, \bibinfo {author} {\bibfnamefont
  {P.}~\bibnamefont {Prettenhofer}}, \bibinfo {author} {\bibfnamefont
  {R.}~\bibnamefont {Weiss}}, \bibinfo {author} {\bibfnamefont
  {V.}~\bibnamefont {Dubourg}}, \bibinfo {author} {\bibfnamefont
  {J.}~\bibnamefont {Vanderplas}}, \bibinfo {author} {\bibfnamefont
  {A.}~\bibnamefont {Passos}}, \bibinfo {author} {\bibfnamefont
  {D.}~\bibnamefont {Cournapeau}}, \bibinfo {author} {\bibfnamefont
  {M.}~\bibnamefont {Brucher}}, \bibinfo {author} {\bibfnamefont
  {M.}~\bibnamefont {Perrot}},\ and\ \bibinfo {author} {\bibfnamefont
  {{\'{E}}.}~\bibnamefont {Duchesnay}},\ }\bibfield  {title} {\bibinfo {title}
  {{Scikit-learn: Machine Learning in Python}},\ }\href@noop {} {\bibfield
  {journal} {\bibinfo  {journal} {J. Mach. Learn. Res.}\ }\textbf {\bibinfo
  {volume} {12}},\ \bibinfo {pages} {2825} (\bibinfo {year}
  {2011})}\BibitemShut {NoStop}%
\bibitem [{\citenamefont {Lei}\ and\ \citenamefont
  {Montoya}(2023)}]{lei2023gmp}%
  \BibitemOpen
  \bibfield  {author} {\bibinfo {author} {\bibfnamefont {X.}~\bibnamefont
  {Lei}}\ and\ \bibinfo {author} {\bibfnamefont {J.}~\bibnamefont {Montoya}},\
  }\bibfield  {title} {\bibinfo {title} {Gmp-featurizer: A parallelized python
  package for efficiently computing the gaussian multipole features of atomic
  systems},\ }\href@noop {} {\bibfield  {journal} {\bibinfo  {journal} {Journal
  of Open Source Software}\ }\textbf {\bibinfo {volume} {8}},\ \bibinfo {pages}
  {5476} (\bibinfo {year} {2023})}\BibitemShut {NoStop}%
\bibitem [{\citenamefont {Kresse}\ and\ \citenamefont
  {Joubert}(1999)}]{kresse1999ultrasoft}%
  \BibitemOpen
  \bibfield  {author} {\bibinfo {author} {\bibfnamefont {G.}~\bibnamefont
  {Kresse}}\ and\ \bibinfo {author} {\bibfnamefont {D.}~\bibnamefont
  {Joubert}},\ }\bibfield  {title} {\bibinfo {title} {From ultrasoft
  pseudopotentials to the projector augmented-wave method},\ }\href@noop {}
  {\bibfield  {journal} {\bibinfo  {journal} {Physical review b}\ }\textbf
  {\bibinfo {volume} {59}},\ \bibinfo {pages} {1758} (\bibinfo {year}
  {1999})}\BibitemShut {NoStop}%
\bibitem [{\citenamefont {Vinyals}\ \emph {et~al.}(2015)\citenamefont
  {Vinyals}, \citenamefont {Bengio},\ and\ \citenamefont
  {Kudlur}}]{vinyals2015order}%
  \BibitemOpen
  \bibfield  {author} {\bibinfo {author} {\bibfnamefont {O.}~\bibnamefont
  {Vinyals}}, \bibinfo {author} {\bibfnamefont {S.}~\bibnamefont {Bengio}},\
  and\ \bibinfo {author} {\bibfnamefont {M.}~\bibnamefont {Kudlur}},\
  }\bibfield  {title} {\bibinfo {title} {Order matters: Sequence to sequence
  for sets},\ }\href@noop {} {\bibfield  {journal} {\bibinfo  {journal} {arXiv
  preprint arXiv:1511.06391}\ } (\bibinfo {year} {2015})}\BibitemShut {NoStop}%
\bibitem [{\citenamefont {Virtanen}\ \emph {et~al.}(2020)\citenamefont
  {Virtanen}, \citenamefont {Gommers}, \citenamefont {Oliphant}, \citenamefont
  {Haberland}, \citenamefont {Reddy}, \citenamefont {Cournapeau}, \citenamefont
  {Burovski}, \citenamefont {Peterson}, \citenamefont {Weckesser},
  \citenamefont {Bright} \emph {et~al.}}]{virtanen2020scipy}%
  \BibitemOpen
  \bibfield  {author} {\bibinfo {author} {\bibfnamefont {P.}~\bibnamefont
  {Virtanen}}, \bibinfo {author} {\bibfnamefont {R.}~\bibnamefont {Gommers}},
  \bibinfo {author} {\bibfnamefont {T.~E.}\ \bibnamefont {Oliphant}}, \bibinfo
  {author} {\bibfnamefont {M.}~\bibnamefont {Haberland}}, \bibinfo {author}
  {\bibfnamefont {T.}~\bibnamefont {Reddy}}, \bibinfo {author} {\bibfnamefont
  {D.}~\bibnamefont {Cournapeau}}, \bibinfo {author} {\bibfnamefont
  {E.}~\bibnamefont {Burovski}}, \bibinfo {author} {\bibfnamefont
  {P.}~\bibnamefont {Peterson}}, \bibinfo {author} {\bibfnamefont
  {W.}~\bibnamefont {Weckesser}}, \bibinfo {author} {\bibfnamefont
  {J.}~\bibnamefont {Bright}}, \emph {et~al.},\ }\bibfield  {title} {\bibinfo
  {title} {Scipy 1.0: fundamental algorithms for scientific computing in
  python},\ }\href@noop {} {\bibfield  {journal} {\bibinfo  {journal} {Nature
  methods}\ }\textbf {\bibinfo {volume} {17}},\ \bibinfo {pages} {261}
  (\bibinfo {year} {2020})}\BibitemShut {NoStop}%
\end{thebibliography}%

\section*{Conflicts of interest}
There are no conflicts of interest to declare. 

Certain commercial products or company names are identified here to describe our study adequately. Such identification is not intended to imply recommendation or endorsement by the National Institute of Standards and Technology, nor is it intended to imply that the products or names identified are necessarily the best available for the purpose.

\section*{Acknowledgments}
This research was undertaken thanks in part to funding provided to the University of Toronto's Acceleration Consortium from the Canada First Research Excellence Fund (Grant number: CFREF-2022-00042). Computational resources were provided by the Calcul Quebec, Westgrid, and Compute Ontario consortia in the Digital Research Alliance of Canada, and the Acceleration Consortium at the University of Toronto. We thank Rhys Goodall for constructive comments and discussion.

\section*{Author contributions}
K.L. and J.H.-S. conceived the idea. K.L. designed the project, performed the investigation, and wrote the manuscript. A.N.R., X.L., and D.P. assisted in the investigation. J.H.-S. supervised the project. All authors contributed to the discussion and revision.
\end{document}

\title{Supplemental Material for: Probing generalizability in machine learning for materials science}

\author{Kangming Li}
\email[]{kangming.li@utoronto.ca}
\affiliation{Department of Materials Science and Engineering, University of Toronto, 27 King’s College Cir, Toronto, ON, Canada.}

\author{Andre Niyongabo Rubungo}
\affiliation{Department of Computer Science, Princeton University.}

\author{Xiangyun Lei}
\affiliation{Toyota Research Institute, 4440 El Camino Real, Los Altos, California 94022, USA.
}

\author{Daniel Persaud}
\affiliation{Department of Materials Science and Engineering, University of Toronto, 27 King’s College Cir, Toronto, ON, Canada.}

\author{Kamal Choudhary}
\affiliation{Material Measurement Laboratory, National Institute of Standards and Technology, 100 Bureau Dr, Gaithersburg, MD, USA.}

\author{Brian DeCost}
\affiliation{Material Measurement Laboratory, National Institute of Standards and Technology, 100 Bureau Dr, Gaithersburg, MD, USA.}

\author{Adji Bousso Dieng}
\affiliation{Department of Computer Science, Princeton University.}

\author{Jason Hattrick-Simpers}
\email[]{jason.hattrick.simpers@utoronto.ca}
\affiliation{Department of Materials Science and Engineering, University of Toronto, 27 King’s College Cir, Toronto, ON, Canada.}
\affiliation{Acceleration Consortium, University of Toronto, 27 King’s College Cir, Toronto, ON, Canada.}
\affiliation{Vector Institute for Artificial Intelligence, 661 University Ave, Toronto, ON, Canada.}
\affiliation{Schwartz Reisman Institute for Technology and Society, 101 College St, Toronto, ON, Canada.} 

\maketitle

\tableofcontents

\section{Chemistry-based OOD generalization}

\subsection{Leave one element out tasks}

\subsubsection{Data distribution across periodic table}

\begin{figure}
    \centering
\foreach \n in {jarvis22,mp21,oqmd21}{
    \includegraphics[width=0.49\linewidth]{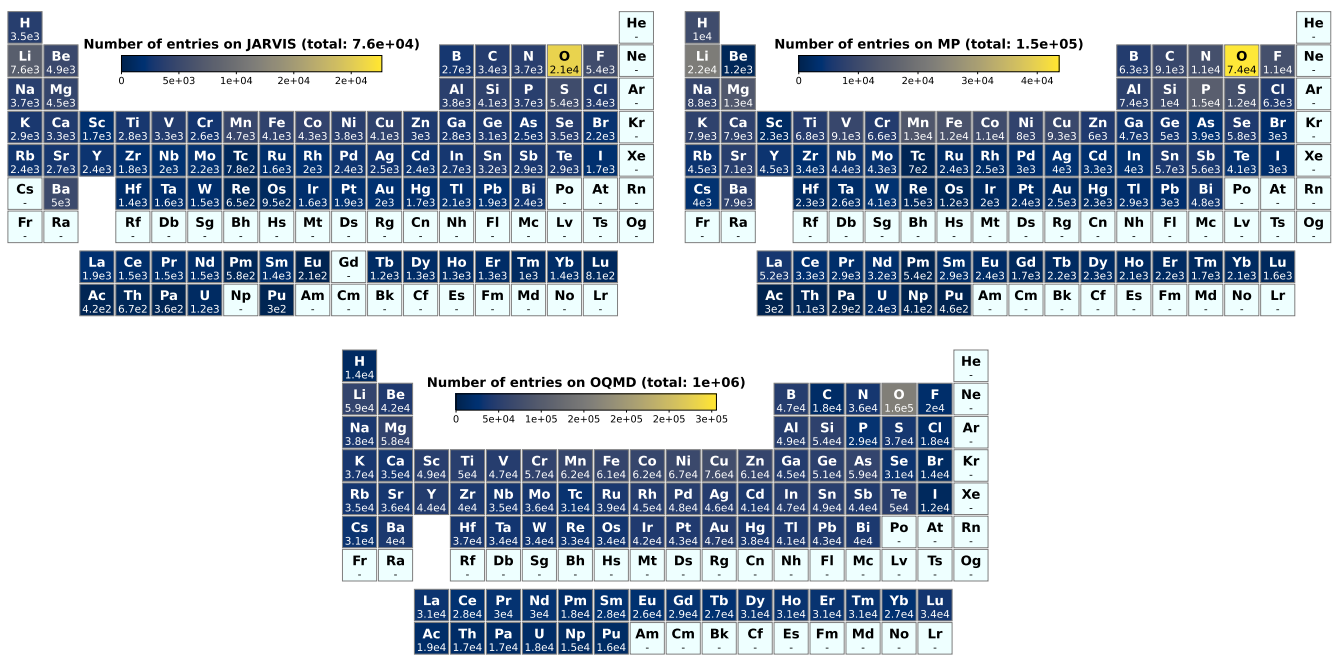}
}
    \caption{Distribution of formation energy data for $X$-containing structures in JARVIS, MP, and OQMD. The total number of entries is indicated in the colorbar title.}
    \label{fig:n_entries e_form}
\end{figure}

\FloatBarrier
\subsubsection{OOD $R^2$ for energy prediction across periodic table}

\begin{figure}
    \centering
\foreach \n in {jarvis22,mp21,oqmd21}{
    \includegraphics[width=0.49\linewidth]{Suppl_Fig/ptable_e_form_\n_rf_r2.pdf}
    \includegraphics[width=0.49\linewidth]{Suppl_Fig/ptable_e_form_\n_xgb_r2.pdf}
}
    \caption{Random forest and XGBoost performance.}
    \label{fig:r2 tree}
\end{figure}

\begin{figure}
    \centering
\foreach \n in {jarvis22,mp21}{
    \includegraphics[width=0.49\linewidth]{Suppl_Fig/ptable_e_form_\n_gmp_r2.pdf}
    \includegraphics[width=0.49\linewidth]{Suppl_Fig/ptable_e_form_\n_alignn25_r2.pdf}
}
    \caption{GMP and ALIGNN performance.}
    \label{fig:r2 gmp alignn}
\end{figure}

\begin{figure}
    \centering
    \includegraphics[width=0.49\linewidth]{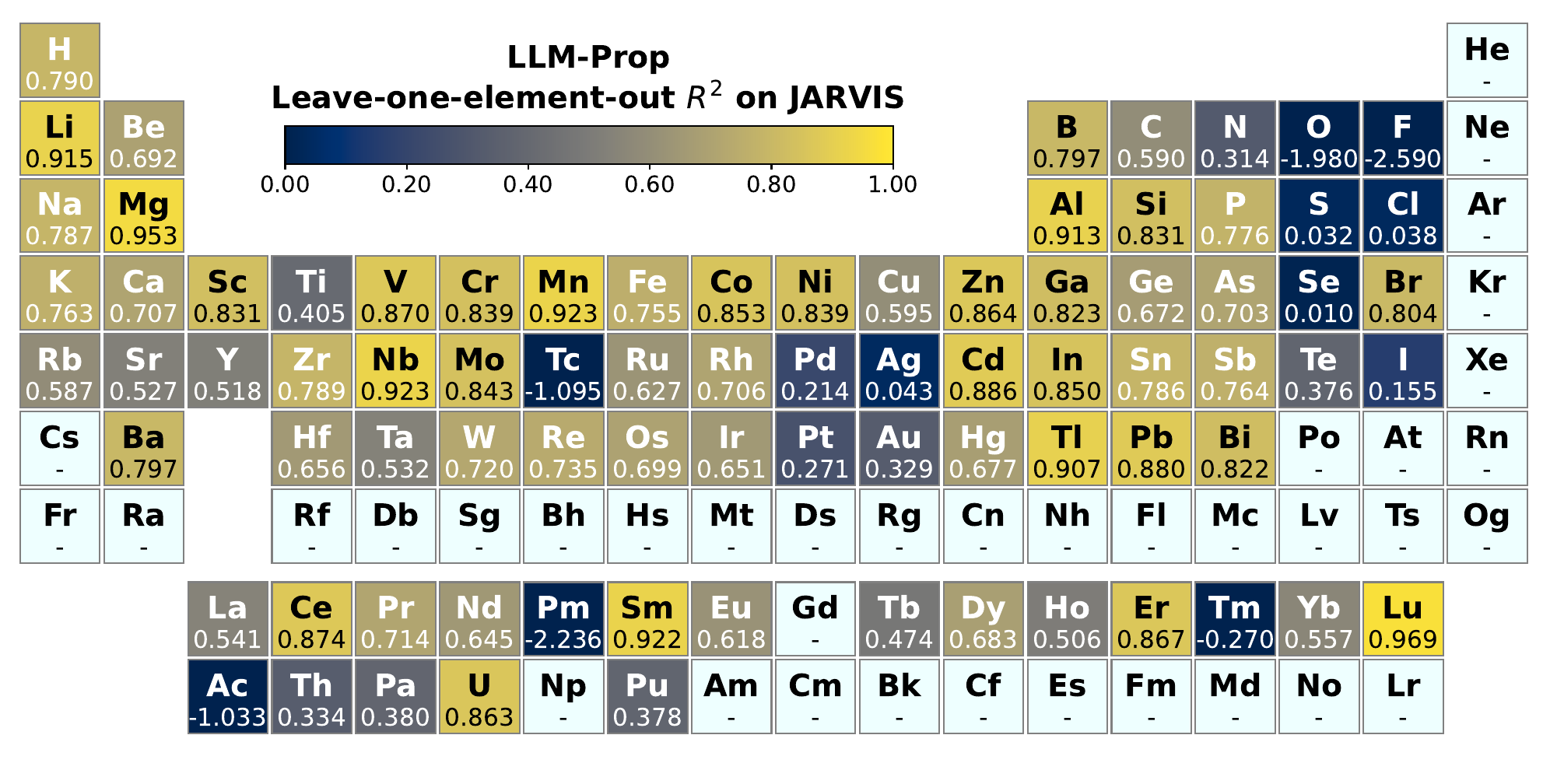}
    \caption{LLM-Prop performance.}
    \label{fig:r2 llm}
\end{figure}

\FloatBarrier
\subsubsection{Parity plots}

% (for elements in the first four periods excluding noble gases) on Materials Project formation energy dataset. 
For all the subplots, the $X$ and $Y$ axis indicate the DFT ground truth and ALIGNN predictions, respectively, and have the same range of -5 to 5 eV/atom

\begin{figure*}
    \centering
\foreach \n in {H,Li,Be,B,C,N,O,F,Na,Mg,Al,Si,P,S,Cl,K,Ca,Sc,Ti,V,Cr,Mn,Fe,Co,Ni,Cu,Zn,Ga,Ge,As,Se,Br}{
    \includegraphics[width=0.19\linewidth]{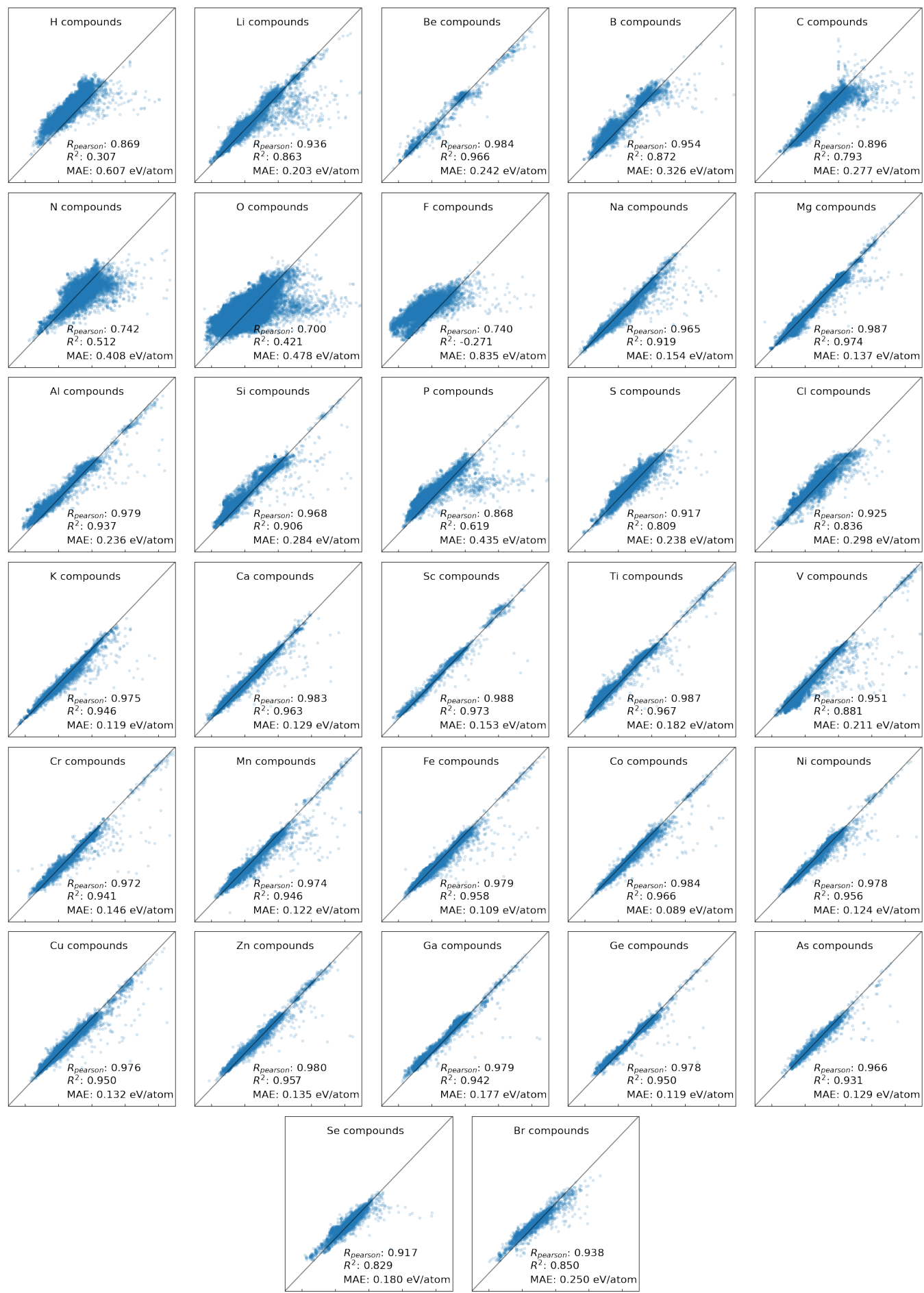}
}
    \caption{Parity plots of random forest predictions on Materials Project formation energy dataset.}
    \label{fig:parity element rf}
\end{figure*}

\begin{figure*}
    \centering
\foreach \n in {H,Li,Be,B,C,N,O,F,Na,Mg,Al,Si,P,S,Cl,K,Ca,Sc,Ti,V,Cr,Mn,Fe,Co,Ni,Cu,Zn,Ga,Ge,As,Se,Br}{
    \includegraphics[width=0.19\linewidth]{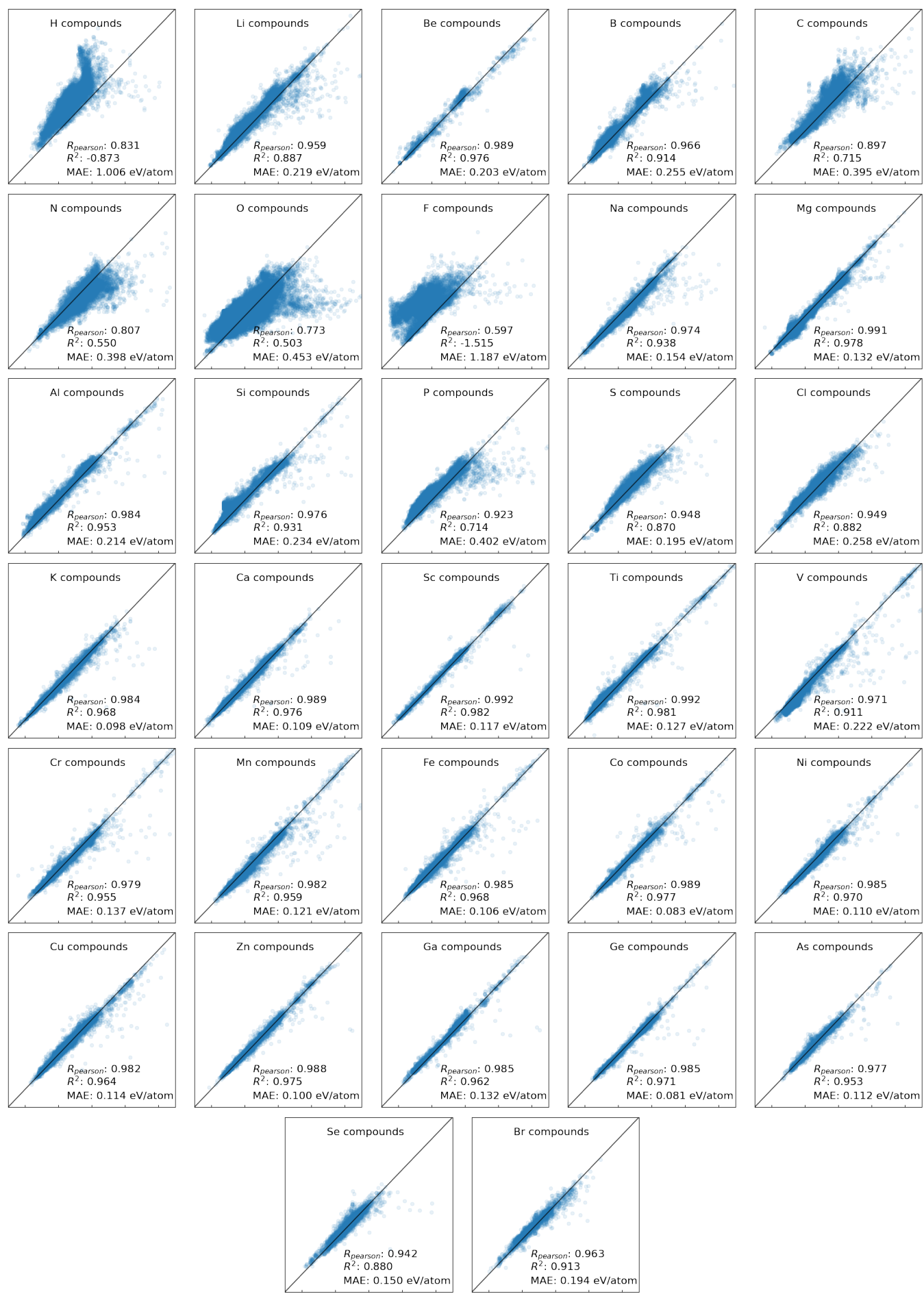}
}
    \caption{Parity plots of XGBoost predictions on Materials Project formation energy dataset.}
    \label{fig:parity element xgb}
\end{figure*}

\begin{figure*}
    \centering
\foreach \n in {H,Li,Be,B,C,N,O,F,Na,Mg,Al,Si,P,S,Cl,K,Ca,Sc,Ti,V,Cr,Mn,Fe,Co,Ni,Cu,Zn,Ga,Ge,As,Se,Br}{
    \includegraphics[width=0.19\linewidth]{Suppl_Fig/parity_plot_e_form_mp21_gmp_elements_\n.png}
}
    \caption{Parity plots of GMP predictions on Materials Project formation energy dataset.}
    \label{fig:parity element gmp}
\end{figure*}

\begin{figure*}
    \centering
\foreach \n in {H,Li,Be,B,C,N,O,F,Na,Mg,Al,Si,P,S,Cl,K,Ca,Sc,Ti,V,Cr,Mn,Fe,Co,Ni,Cu,Zn,Ga,Ge,As,Se,Br}{
    \includegraphics[width=0.19\linewidth]{Suppl_Fig/parity_plot_e_form_mp21_alignn25_elements_\n.png}
}
    \caption{Parity plots of ALIGNN predictions on Materials Project formation energy dataset.}
    \label{fig:parity element alignn}
\end{figure*}

\begin{figure*}
    \centering
\foreach \n in {H,F,O,P,I,Li,Pt,Bi,Fe,Rb}{
    \includegraphics[width=0.19\linewidth]{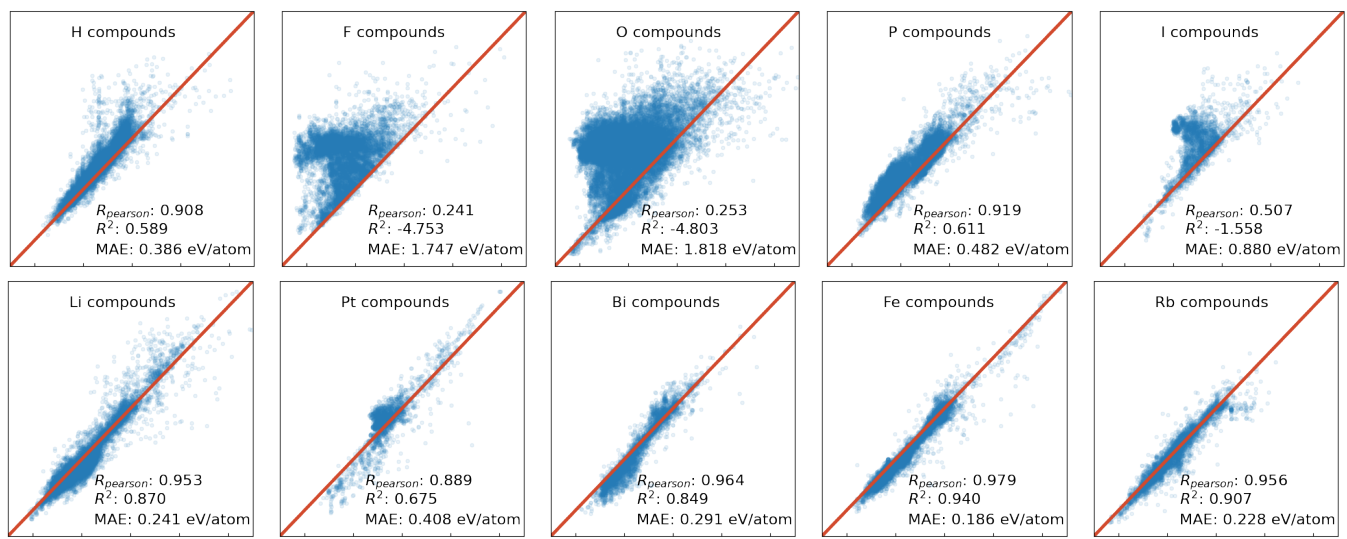}
}
    \caption{Parity plots of LLM-Prop predictions on Materials Project formation energy dataset.}
    \label{fig:parity element llm}
\end{figure*}

\FloatBarrier
\subsection{Leave one period out tasks}

\subsubsection{Parity plots}

\begin{figure*}
    \centering
\foreach \n in {2,...,7}{
    \includegraphics[width=0.32\linewidth]{Suppl_Fig/parity_plot_e_form_mp21_alignn25_period_\n.png}
}
    \caption{ALIGNN model performance. For all the subplots, the X and Y axis indicate the DFT ground truth and ALIGNN predictions, respectively, and have the same range of -5 to 5 eV/atom.}
    \label{fig:parity period alignn}
\end{figure*}

\FloatBarrier
\subsection{Leave one group out tasks}
\subsubsection{Parity plots}

\begin{figure*}
    \centering
\foreach \n in {1,...,17}{
    \includegraphics[width=0.24\linewidth]{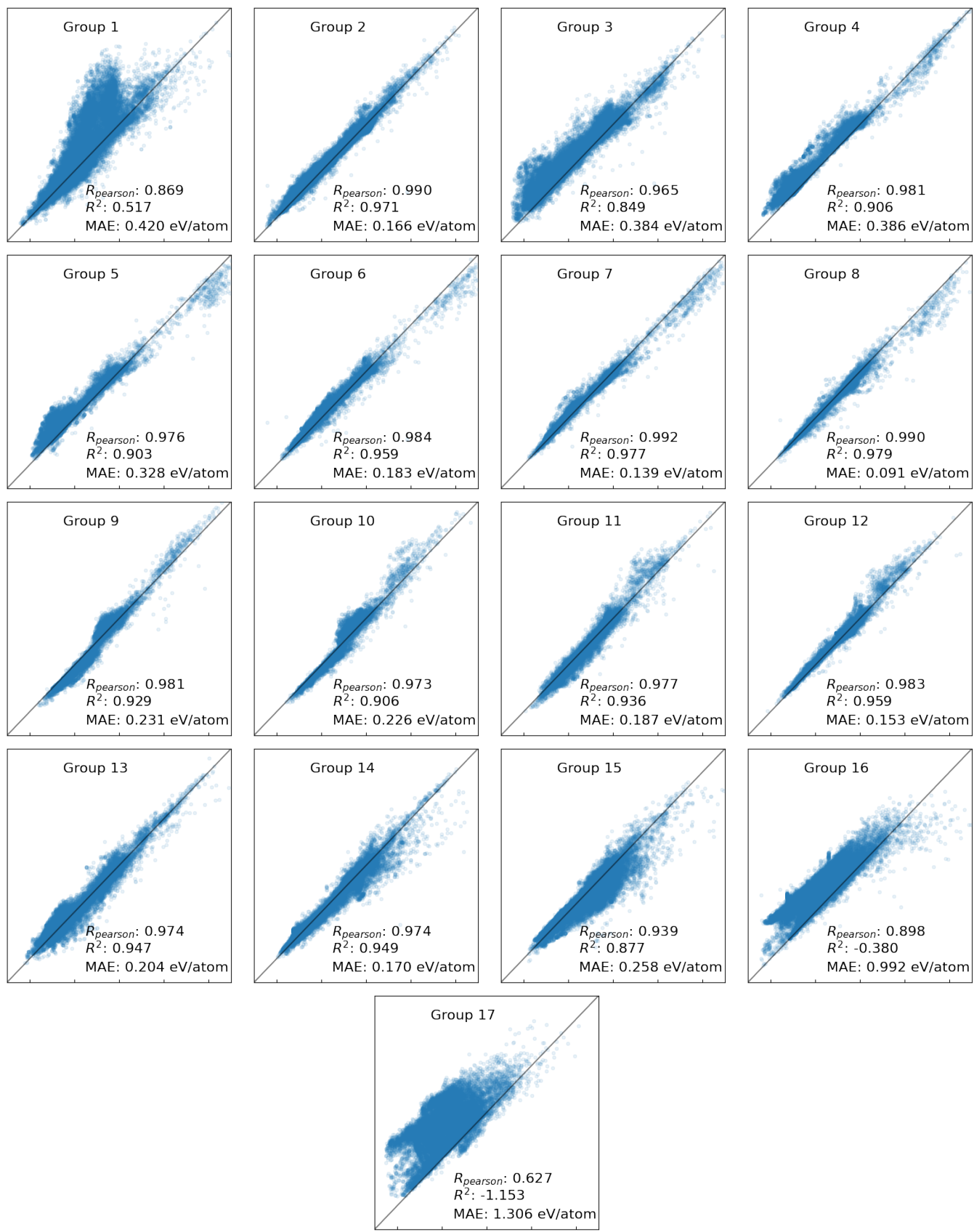}
}
    \caption{ALIGNN model performance. For all the subplots, the X and Y axis indicate the DFT ground truth and ALIGNN predictions, respectively, and have the same range of -5 to 5 eV/atom.}
    \label{fig:parity group alignn}
\end{figure*}

\FloatBarrier
\section{Structure-based OOD generalization}

\subsection{Leave one prototype-like out tasks}

\subsubsection{Parity plots}

\begin{figure*}
    \centering
\foreach \n in {Heusler,Indium,Hydrophilite,Perovskite, Spinel,Caswellsilverite}{
    \includegraphics[width=0.25\linewidth]{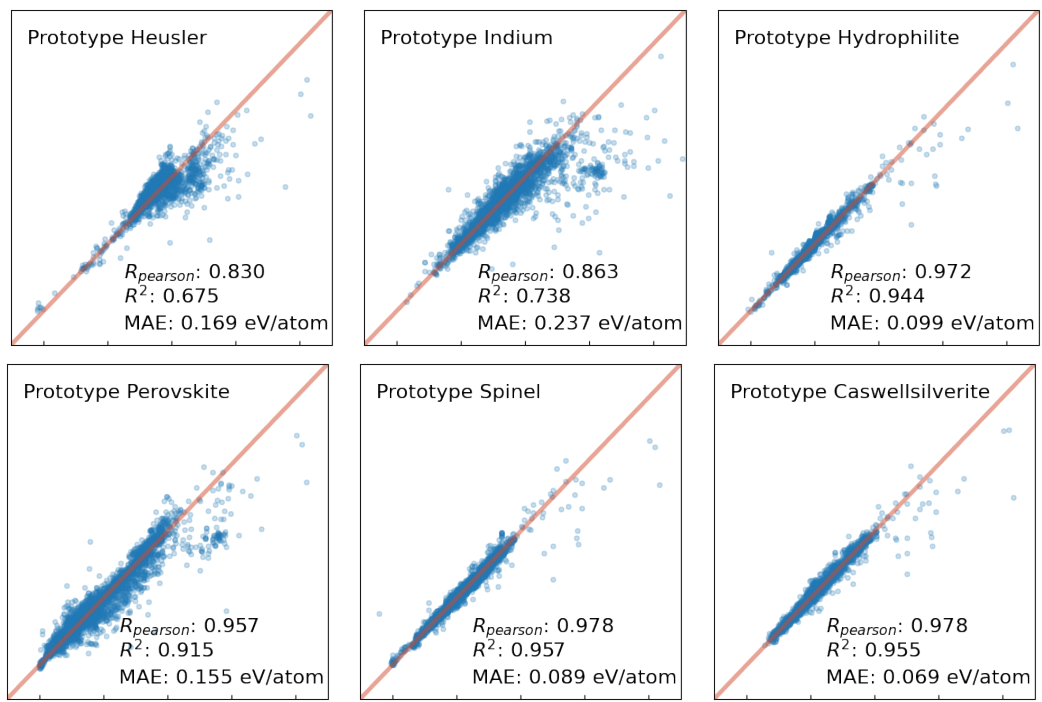}
}
    \caption{Parity plots of XGB predictions on Materials Project formation energy dataset.}
    \label{fig:parity prototype xgb}
\end{figure*}

\begin{figure*}
    \centering
\foreach \n in {Heusler, Indium,Hydrophilite,Perovskite, Spinel,Caswellsilverite}{
    \includegraphics[width=0.25\linewidth]{Suppl_Fig/parity_plot_e_form_mp21_gmp_prototype_\n.png}
}
    \caption{Parity plots of GMP predictions on Materials Project formation energy dataset.}
    \label{fig:parity prototype gmp}
\end{figure*}

\begin{figure*}
    \centering
\foreach \n in {Heusler, Indium,Hydrophilite,Perovskite, Spinel,Caswellsilverite}{
    \includegraphics[width=0.25\linewidth]{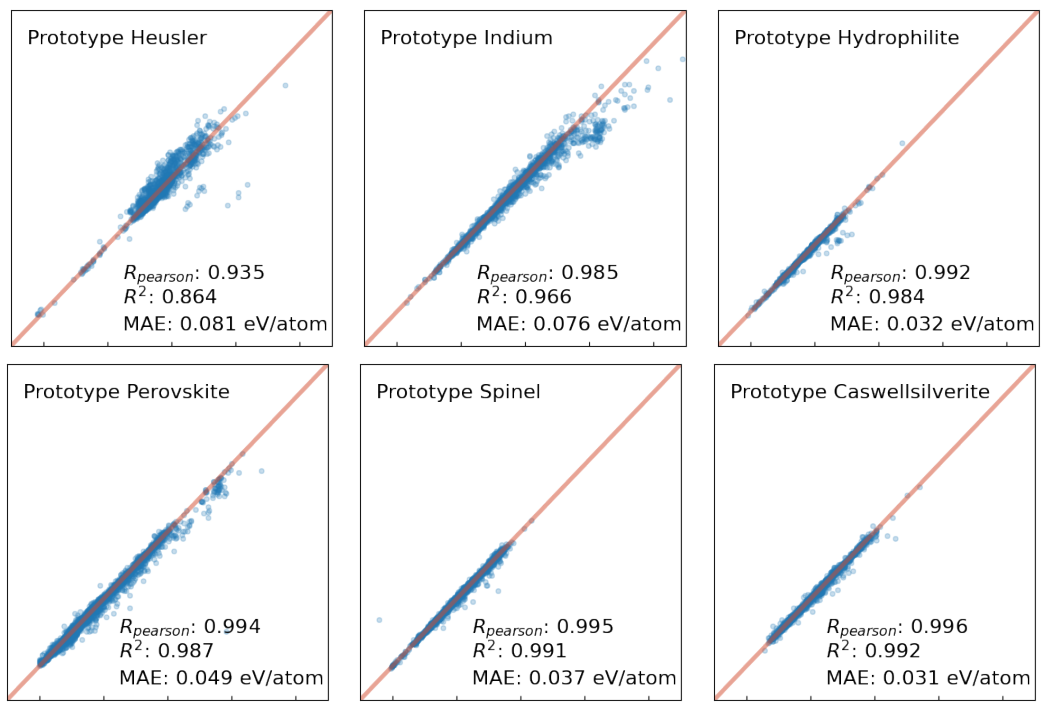}
}
    \caption{Parity plots of ALIGNN predictions on Materials Project formation energy dataset.}
    \label{fig:parity prototype alignn25}
\end{figure*}

\FloatBarrier
\section{Learning curves for out-of-distribution generalization}

\subsection{MAE versus training time}

\begin{figure*}
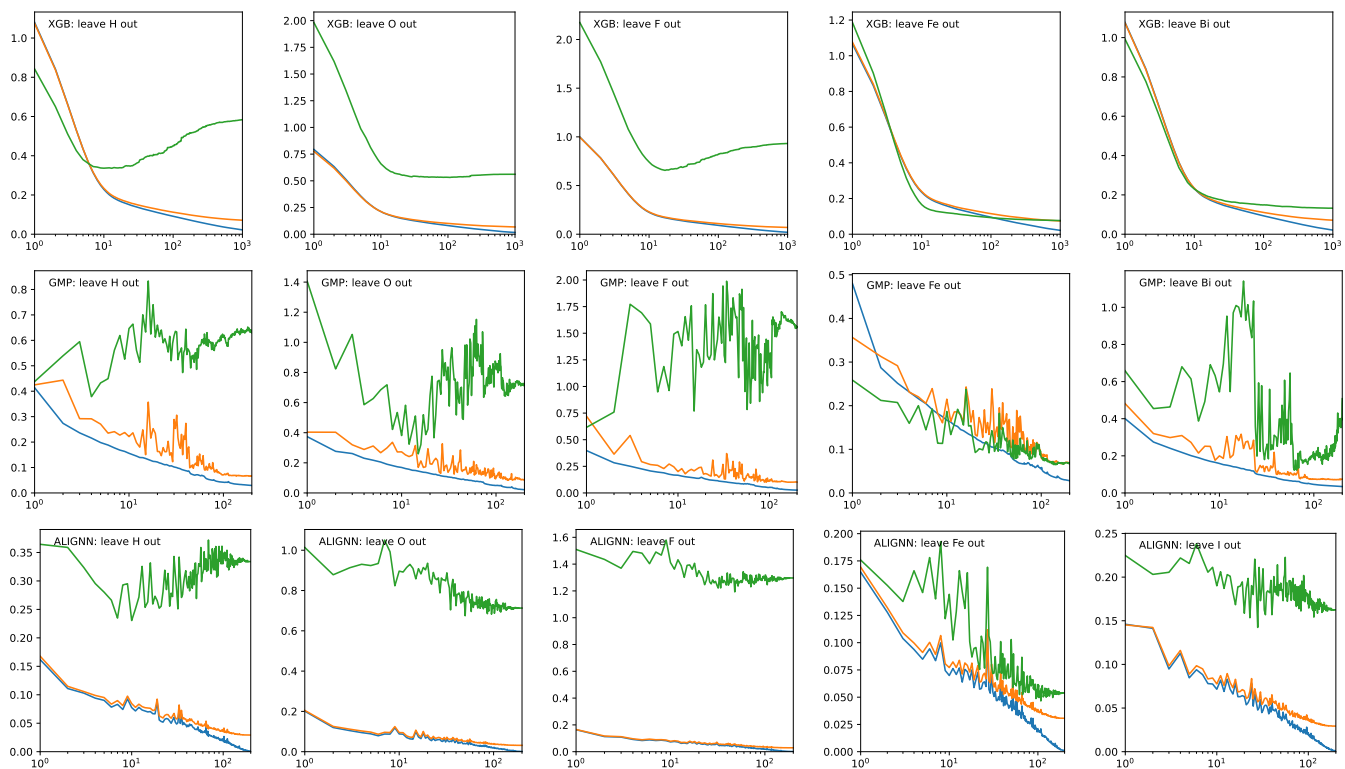

    \centering
    
% \foreach \n in {H,Li,O,F,Mn,Fe,Ba,Bi}{
%     \includegraphics[width=0.24\linewidth]{Suppl_Fig/learning_curves/learning_curve_jarvis22_elements_\n_e_form_xgb.pdf}
% }
% \foreach \n in {H,C,O,F,Si,Fe,Ba,Bi}{
%     \includegraphics[width=0.24\linewidth]{Suppl_Fig/learning_curves/learning_curve_jarvis22_elements_\n_e_form_gmp.pdf}
% }
% \foreach \n in {H,Li,O,F,Mg,Fe,N,I}{
%     \includegraphics[width=0.24\linewidth]{Suppl_Fig/learning_curves/learning_curve_jarvis22_elements_\n_e_form_alignn200.pdf}
% }

\foreach \n in {H,O,F,Fe,Bi}{
    \includegraphics[width=0.19\linewidth]{Suppl_Fig/learning_curves/learning_curve_jarvis22_elements_\n_e_form_xgb.pdf}
}
\foreach \n in {H,O,F,Fe,Bi}{
    \includegraphics[width=0.19\linewidth]{Suppl_Fig/learning_curves/learning_curve_jarvis22_elements_\n_e_form_gmp.pdf}
}
\foreach \n in {H,O,F,Fe,I}{
    \includegraphics[width=0.19\linewidth]{Suppl_Fig/learning_curves/learning_curve_jarvis22_elements_\n_e_form_alignn200.pdf}
}
    \caption{Learning curves for XGBoost, GMP, and ALIGNN models. The $X$ axis is the number of boosting rounds (XGBoost) or epochs (GMP and ALIGNN). The Y axis is the MAE in eV/atom. The lines in blue, orange, and green represent the training loss, in-distribution test loss, and out-of-distribution test loss, respectively.}
    \label{fig:jarvis22 training time}
\end{figure*}

\FloatBarrier
\subsection{MAE versus training set size}

\begin{figure*}
    \centering
    \includegraphics[width=0.4\linewidth]{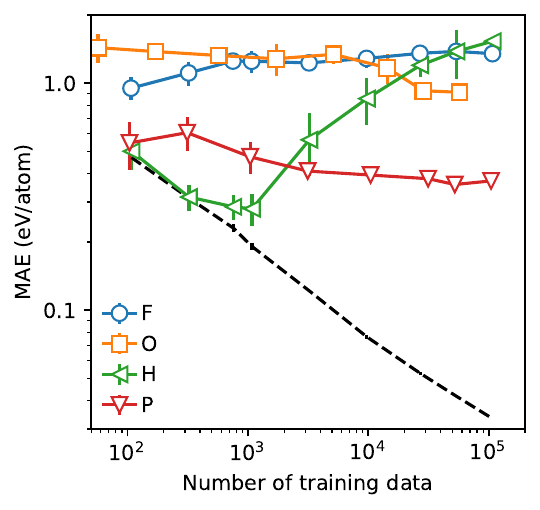}
    \includegraphics[width=0.4\linewidth]{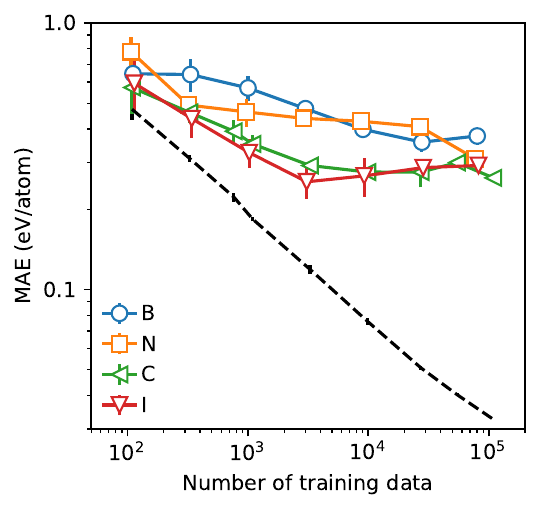}
    \includegraphics[width=0.4\linewidth]{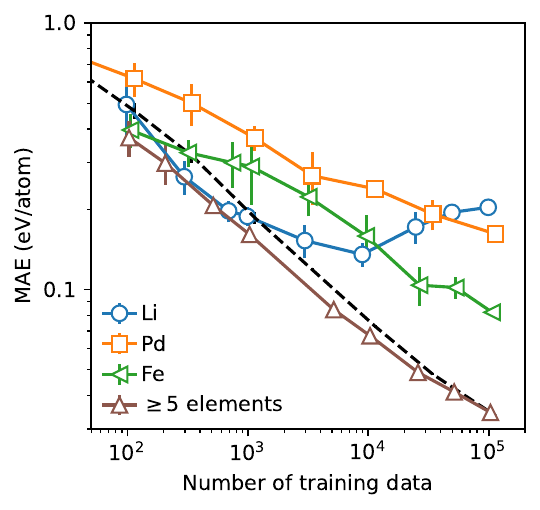}
    \caption{ALIGNN MAE versus training set size for Materials Project.}
    \label{fig:training set size mp}
\end{figure*}

\begin{figure*}
    \centering
    \includegraphics[width=0.4\linewidth]{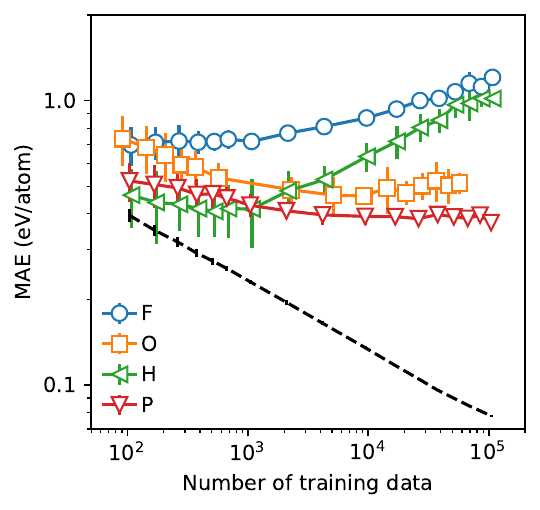}
    \includegraphics[width=0.4\linewidth]{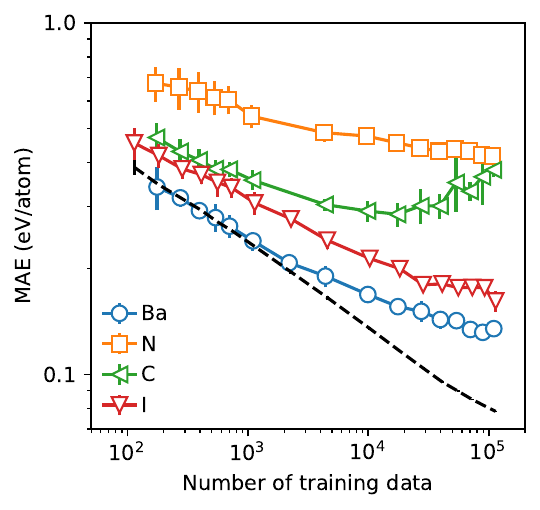}
    \includegraphics[width=0.4\linewidth]{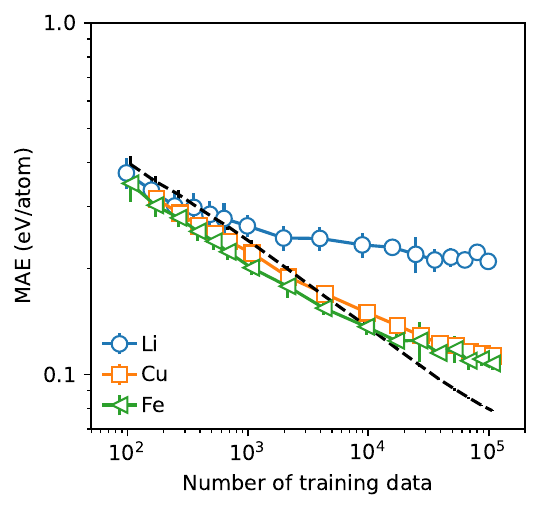}
    \caption{XGB MAE versus training set size for Materials Project.}
    \label{fig:training set size mp xgb}
\end{figure*}

\begin{figure*}
    \centering    
\foreach \n in {H,F,Li,C,Fe}{
    \includegraphics[width=0.3\linewidth]{Suppl_Fig/ood_alignn25_mp21_e_form_elements_\n_err.pdf}
}
    \includegraphics[width=0.3\linewidth]{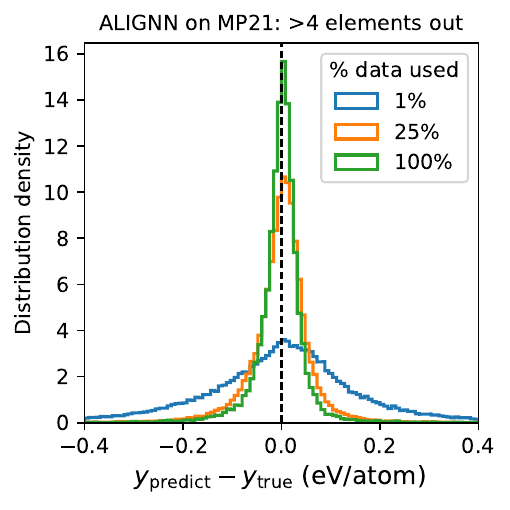}
    \caption{Distribution of ALIGNN prediction errors as functions of Materials Project training set sizes for tasks in Fig.~\ref{fig:training set size mp}.}
    \label{fig:mp training set size}
\end{figure*}

\begin{figure*}
    \centering
    \includegraphics[width=0.4\linewidth]{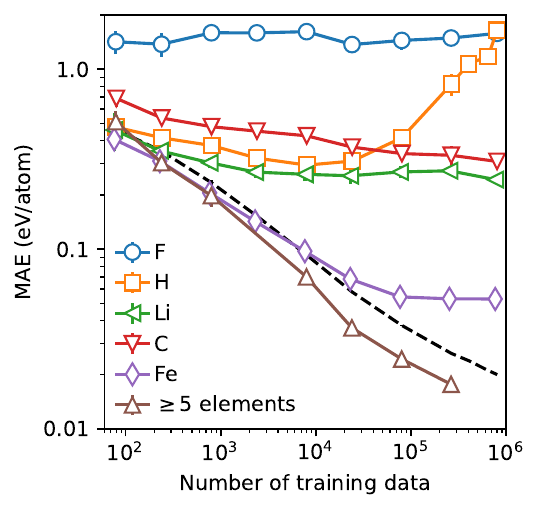}
    \caption{ALIGNN MAE versus training set size for OQMD.}
    \label{fig:training set size oqmd}
\end{figure*}

\begin{figure*}
    \centering
    
\foreach \n in {H,F,Li,C,Fe}{
    \includegraphics[width=0.3\linewidth]{Suppl_Fig/ood_alignn25_oqmd21_e_form_elements_\n_err.pdf}
}
    \includegraphics[width=0.3\linewidth]{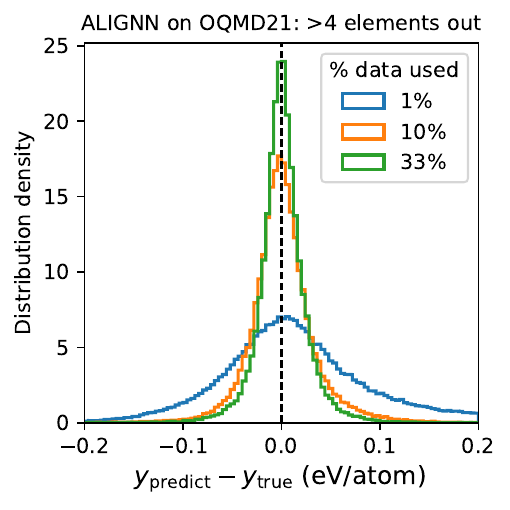}
    \caption{Distribution of ALIGNN prediction errors as functions of OQMD training set sizes for tasks in Fig.~\ref{fig:training set size oqmd}.}
    \label{fig:oqmd21 training set size}
\end{figure*}

\begin{figure*}
    \centering
    \includegraphics[width=0.4\linewidth]{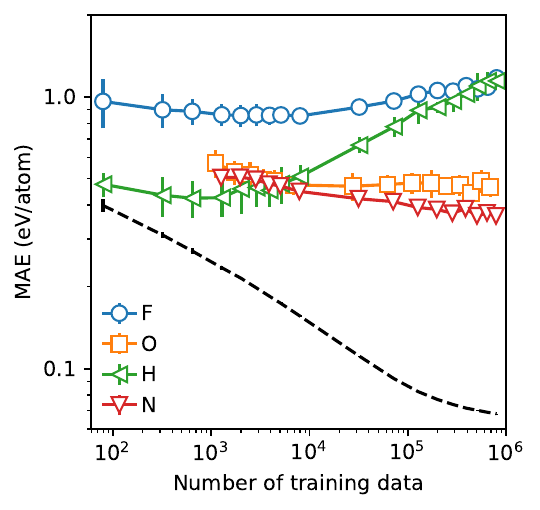}
    \includegraphics[width=0.4\linewidth]{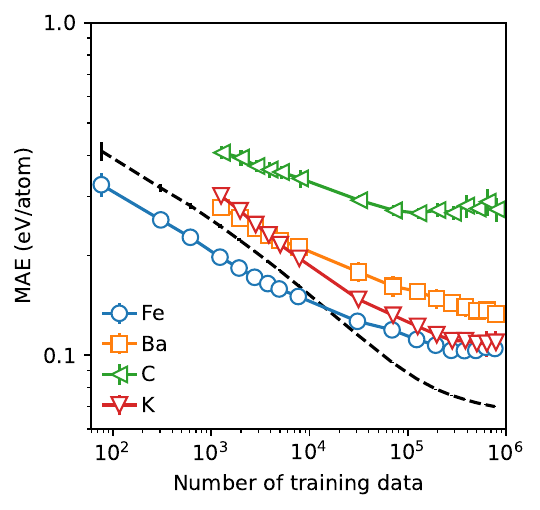}
    \includegraphics[width=0.4\linewidth]{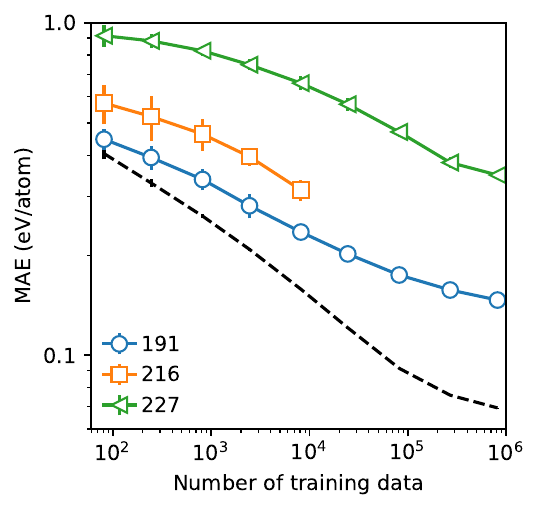}
    \caption{XGB MAE versus training set size for OQMD. The first two plots are for the leave-one-element-out tasks and the third for the leave-one-space-group-out tasks.}
    \label{fig:training set size oqmd xgb}
\end{figure*}